\documentclass[twocol]{ametsoc}
\usepackage[mathscr]{eucal}

\journal{jpo}

\bibpunct{(}{)}{;}{a}{}{,}

\newcommand{\beq}{\begin{equation}}
\newcommand{\eeq}{\end{equation}}

\newcommand{\defn}{\ensuremath{\stackrel{\mathrm{def}}{=}}}

\newcommand{\com}{\, ,}

\newcommand{\ybjp}{YBJ$^{+}$}

\newcommand{\p}{\partial}

\newcommand{\laa}{\left \langle}
\newcommand{\raa}{\right \rangle}

\newcommand{\lap}{\triangle}

\newcommand{\ii}{\mathrm{i}}
\newcommand{\LL}{\mathsf{L}}
\newcommand{\LLp}{\mathsf{L}^{\!\!\scriptscriptstyle+}\!}

\newcommand{\halfi}{\tfrac{\ii}{2}}

\newcommand{\Bu}{\mbox{\textit{Bu}}} 

\usepackage{float}
\usepackage{mwe,tikz}\usepackage[percent]{overpic}

\title{Penetration of wind-generated near-inertial waves into a turbulent ocean}

\authors{Olivier Asselin\correspondingauthor{Olivier Asselin, Keck 254, Scripps Institution of Oceanography, University of California San Diego, La Jolla, CA 90293-0213, USA} and William R. Young}

\affiliation{Scripps Institution of Oceanography, University of California San Diego, La Jolla, USA} 

\email{oasselin@ucsd.edu}

\abstract{An idealized storm scenario is examined in which a wind-generated inertial wave interacts with a turbulent baroclinic quasi-geostrophic flow. The flow is initialized by spinning up a Eady model with a realistic stratification profile. The storm is modeled as an initial value problem for a mixed-layer confined, horizontally-uniform inertial oscillation. The primordial inertial oscillation then evolves under the effects of advection, refraction, dispersion and dissipation. Waves feedback onto the flow by modifying its potential vorticity. In the first few days, refraction dominates and wave energy is attracted (repelled) by regions of negative (positive) vorticity. Wave energy is subsequently drained down anticyclonic pipes. This drainage halts as wave energy encounters weakening vorticity. After a week or two, wave energy accumulates at the bottom of negative vorticity features, \emph{i.e.} along filamentary structures at shallow depths and in larger anticyclones at greater depths. Wave feedback tends to weaken vortices and thus slow down wave penetration. This effect, however, is found to be weak even for vigorous storms.}

\begin{document}

\maketitle
 
\section{Introduction}

Near-inertial waves  comprise half of the energy and most of the vertical shear of the ocean internal wave spectrum \citep{ferrari2009}. These waves most strikingly manifest as near-circular surface drifter orbits \citep{dasaro1995} and as a ubiquitous spectral peak in moored current profiler records \citep{webster1968,fu1981,alford2016review}. Because of their large vertical shear, near-inertial waves are thought to be major drivers of upper-ocean mixing \citep{lueck1986dissipation,kunze1995}.


Near-inertial waves originate in the ocean mixed layer with the 1000km horizontal scale characteristic of atmospheric storms \citep{pollard1980,dasaro1995}. Were these waves to preserve this large initial horizontal scale, there would never be significant penetration --- or contribution to mixing --- below the mixed layer \citep{gill1984}. If the wave frequency is close to the Coriolis frequency $f$, then vertical group velocity is
\beq
c_g^z \approx \Bu \, f/m\com
\label{Gill1}
\eeq
where $m$ is the vertical wavenumber.  The Burger number in \eqref{Gill1}  is $\Bu = (N k/fm)^2$, with $k$ the horizontal wavenumber, and $N$ the buoyancy frequency. With $N/f \sim100$, $k^{-1}\sim 10^6$m and $m^{-1}\sim100$m, the Burger number is of order $10^{-4}$. At mid-latitudes $f \sim 10^{-4}$s$^{-1}$ and the time required to propagate vertically through a distance of  100m  is about three years.  Unless there is a reduction of the horizontal scale from the 1000km generation scale, vertical propagation of near-inertial waves is glacially  slow. 


A main outcome of the Ocean Storms Experiment was observational evidence that the latitudinal variation of the Coriolis frequency --- the  $\beta$-effect ---  leads to a systematic reduction of the  horizontal scale of near-inertial waves  \citep{dasaro1995}. The primordial inertial wave oscillates at different frequencies in its southmost and northmost regions, which results in a de-phasing of the initially-uniform orbits. The  de-phasing increases with time and so produces an ever smaller meridional wavelength, resulting  in significant  vertical propagation of near-inertial waves \citep{dasaro1989}.

These observations, however, were made in a region with weak mesoscale variability \citep{dasaro1995mesoscale}. In fact, atmospheric storm tracks (and thus near-inertial energy) largely coincide with regions of strong mesoscale variability \citep{zhai2005}. In addition to the $\beta$-effect, mesoscale vorticity has long been hypothesized to cause local frequency shifts analog to the $\beta$-effect \citep{mooers1975a,mooers1975b,kunze1985}. These theoretical works predict that vertical vorticity of mesoscale and sub-mesoscale eddies, $\zeta$, shifts the local inertial frequency by $\zeta/2$. Gradients of eddy vorticity are  at least an  order of magnitude larger than the  $\beta$-effect \citep{van1998interactions}.  Thus the $\zeta/2$ frequency shift might be more important than $\beta$ in reducing  horizontal scales and  accelerating vertical propagation. Argo float data shows that both the amplitude of the seasonal cycle of diapycnal mixing and the response to increases in the wind energy flux are larger in regions with  an energetic eddy field \citep{whalen2012,whalen2018}.

Theory predicts that the eddy field imprints its horizontal scale (roughly 10 to 100 km) onto the wave field through refraction. However, early theoretical studies  were largely based on the Wentzel-Kramers-Brillouin (WKB) assumption that  the spatial scale of the waves is much less than that of  the mesoscale flow. This scale-separation assumption is strongly violated by freshly generated near-inertial waves. \cite{ybj} (henceforth, YBJ) developed a model that does not rely on the WKB approximation. Instead, YBJ relies on the time-scale separation between the slow mesoscale and fast near-inertial waves and makes use of a phase average to remove the fast inertial oscillation and expose the slow evolution of the back-rotated near-inertial wave velocity. YBJ also predicts that eddy vorticity shifts the local inertial frequency by a factor $\zeta/2$, thereby confirming  the WKB prediction. 

The YBJ model has first been employed to probe the enhanced propagation of near-inertial waves due to simple vorticity distributions \citep{balmforth1998}, the $\beta$-effect \citep{moehlis2001} or both simultaneously \citep{balmforth1999}. \cite{llewellyn1999} used the YBJ model to calculate the trapped near-inertial modes of  a barotropic axisymmetric vortex. \cite{klein2001} and \cite{klein2004} explored the dispersion of near-inertial waves by a fully turbulent barotropic quasi-gestrophic flow. Decomposition of the wave field into vertical normal modes proved useful to distinguish between trapping and dispersive regimes. \cite{danioux2008propagation} employed a primitive-equation model to study the three-dimensional propagation of wind-generated near-inertial waves. The behavior of shear-containing modes was successfully captured by the YBJ model. These authors also reported a deep maximum of vertical velocity with frequency $2f$ which they further investigated in subsequent works \citep{danioux2008resonance,danioux2011emergence}; see \cite{wy2016} for a model that includes the nonlinearly generated $2f$-harmonic.

The original YBJ model describes the passive distortion of near-inertial waves due to advection and refraction by a balanced flow. \cite{xv} derived a fully coupled model in which strong near-inertial waves may feedback onto the  balanced flow  by modifying its potential vorticity; see also \cite{wagner2015,wy2016} and \cite{salmon2016}. 
\cite{cesar} employed this model to investigate energy exchanges between near-inertial waves and barotropic flows. In absence of flow vertical shear, waves have a fixed vertical wavelength and the wave capture mechanism of \cite{bm2005} does not function. Instead, waves  propagate out of straining regions and this wave escape prevents efficient  transfer of energy.

In this paper we explore how turbulent eddies facilitate the penetration of wind-generated energy into the ocean interior. We consider an idealized storm scenario in which an inertial wave initially confined to the mixed layer interacts with turbulent baroclinic quasi-geostrophic eddies. These eddies are initialized by spinning up a Eady model  with a realistic stratification profile (section \ref{sec:problem}). The subsequent coupled evolution of quasi-geostrophic eddies and near-inertial waves is represented with a novel QG-NIW model (section \ref{sec:model}). Wave evolution is governed by the \ybjp{} model \citep{ybjp}, a higher-order and numerically-docile version of the  YBJ model. Waves also feedback onto the balanced flow by modifying the potential vorticity \citep{xv}. The simulations reported here are the first integrations of a YBJ-like system coupled with a baroclinic flow. This combination provides a  reduced  description of vertical near-inertial propagation into a quasi-geostrophic flow.  The model captures all relevant  physical processes, with the notable exception of mixing. The primitive equations are more realistic, but the wave-eddy decomposition is then challenging and ambiguous; the QG-NIW model has the advantage of representing waves and eddies with distinct variables so that the wave-eddy decomposition is hardwired.  The QG-NIW model thus provides an ideal platform to visualize both the propagation of waves in the wake of a storm (section \ref{sec:prop}) and the effects of waves on eddies (section \ref{sec:feed}). We assemble our findings into a coherent narrative in section \ref{sec:disc}.

\section{Problem setup} \label{sec:problem}

The numerical experiment has two phases: the flow spin-up and the storm, during which waves are introduced. The spin-up generates a realistic turbulent mesoscale flow using the Eady model of baroclinic instability generalized with a realistic stratification profile. Once the Eady solution reaches statistical stationarity, Eady forcing terms are removed and the storm phase begins. The storm is modeled as an initial value problem for the wave field. The wave model is initialized with a horizontally-uniform inertial oscillation in the mixed layer, which subsequently evolves in the mature geostrophic flow. This section describes the initial condition used for the storm experiment.

\subsection{Phase I: Setting up a turbulent mesoscale field}

A generalized Eady model \citep{eady1949} is employed to generate a realistic mesoscale turbulence. Thus a steady vertically-sheared geostrophic base flow is imposed:
\begin{equation}
\Psi =-U(z)y,  \label{meanflow}
\end{equation}
where  $\Psi$ and $U$ are the streamfunction and zonal velocity of the base-state. Meridional and vertical components of the base-state velocity are zero. An Eady model is characterized by a vanishing base-state potential vorticity gradient. This is  achieved if the vertical shear is proportional to the squared buoyancy frequency: $U'(z) \propto N^2(z)$.
The classic Eady model is the special case with  constant $N^2$ and a linearly sheared base-state velocity. Here instead, we use observations from the Near-Inertial Shear and Kinetic Energy Experiment (NISKINe) pilot cruise to set $N^2$.

\begin{figure}
\centering
\includegraphics[width=19pc]{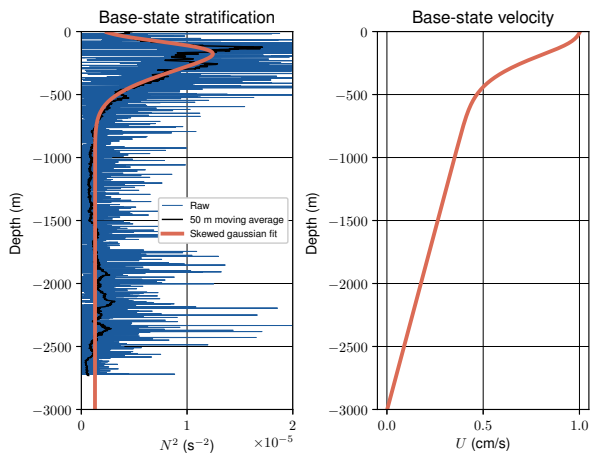}
\caption{Left: Typical raw stratification profile (blue) observed during the NISKINe pilot cruise, along with its 50-meter moving average (black). The orange line is the skewed gaussian fit, \eqref{skewfit}, with parameters outlined in table \ref{N2fit}. Right: Base-state zonal velocity profile ensuring zero potential vorticity gradient with the skewed Gaussian fit for stratification.}
        \label{N2}
\end{figure}

Figure \ref{N2} shows the raw observations (blue), the 50-meter moving average (black) and the fit (red). The observed profile is typical of the month of May in the North Atlantic. There is a relatively well mixed layer in top 50-100m overlaying a themocline that extends down to about 600-700m. Below the thermocline, the abyss has a constant and relatively weak stratification down to the ocean bottom, located at depth of 3km.

We fit the smoothed observed $N^2(z)$ profile with 
\begin{gather}
 N^2_{\mathrm{fit}} =  N^2_0 + N^2_1 \, \text{e}^{-(z-z_0)^2/\sigma^2} \left[ 1 + \text{erf}\left( \frac{\alpha(z-z_0)}{\sigma \sqrt{2}}\right) \right]. \label{skewfit}
\end{gather}
Table \ref{N2fit} displays the values used for the  fit. The abyssal value of stratification, $N^2_0$, is calculated as the average smoothed $N^2$ below 1 km. The right panel of figure \ref{N2} shows the base-state velocity profile $U(z)$. The magnitude of velocity $U(z)$ (and corresponding shear) is a free parameter. We adjust the amplitude of  $U(z)$ so that the  equilibrated eddy field has realistic properties, such as the maximum sea-surface eddy velocity of about 25 cm s$^{-1}$ obtained  from both in-situ observations during the NISKINe cruise, and from geostrophic currents derived from satellite altimetry \citep{bonjean2002diagnostic}. It is remarkable that  this level of eddy energy is produced with the small mean velocity difference  $\Delta U =1$cm s$^{-1}$ indicated in the right panel of figure \ref{N2}.

\begin{table}
	\centering
	\caption{Fitting parameters for $N^2$}
	\footnotesize
	\renewcommand{\arraystretch}{1.3}
	\setlength{\tabcolsep}{11pt}
	\begin{tabular}{r l}
		\hline
		\hline
		Abyssal stratification & $N_0^2 = 1.2927 \times 10^{-6}$ s$^{-2}$ \\
		Gaussian amplitude & $N_1^2 = 6.4532 \times 10^{-6}$ s$^{-2}$ \\
		Gaussian location & $z_0 = -77.1809$ m\\
		Width parameter & $\sigma = 309.6155$ m\\
		Skewness parameter & $\alpha = -5.3384$ \\

		\hline
		\hline
	\end{tabular}  \label{N2fit}
\end{table}

\begin{figure*}[h!]
\centerline{\includegraphics[trim=0 120 0 100, clip, width=0.99\textwidth]{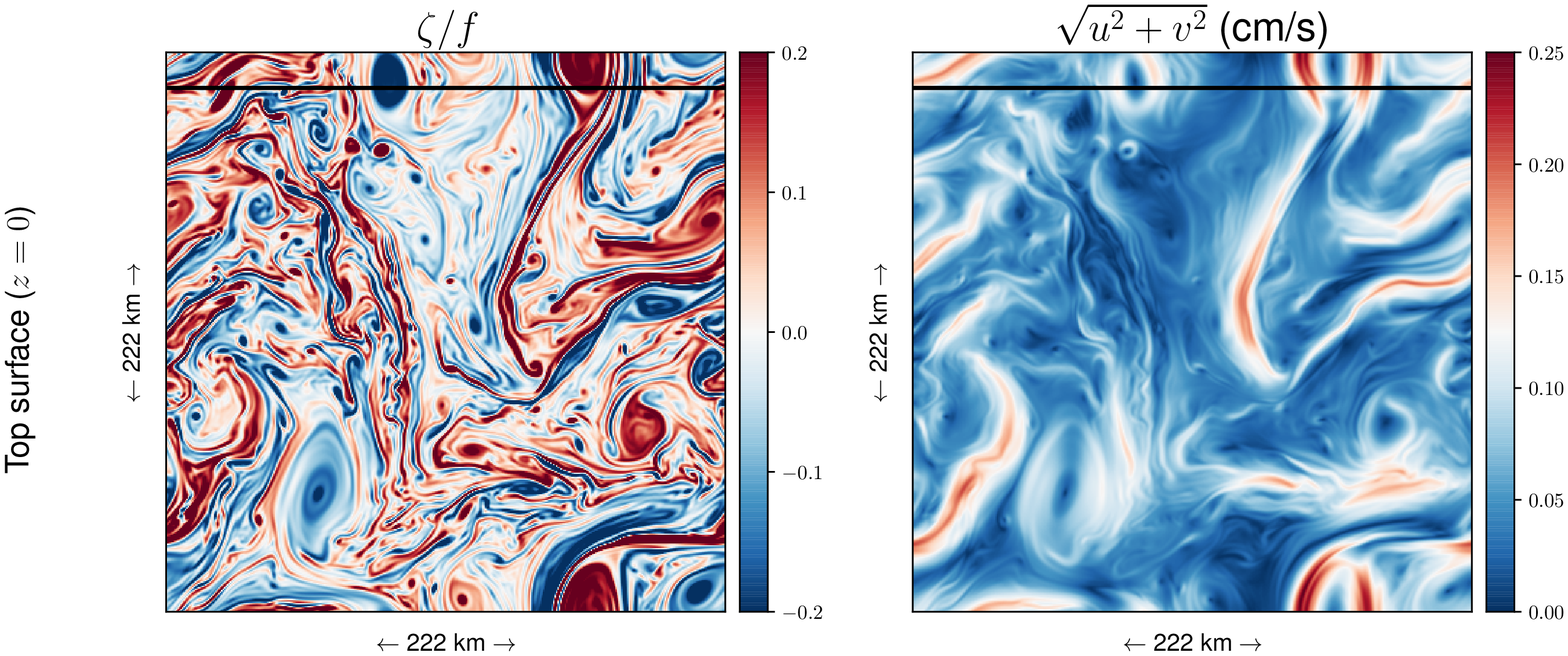}}
\centerline{\includegraphics[trim=0 170 0 190, clip, width=0.99\textwidth]{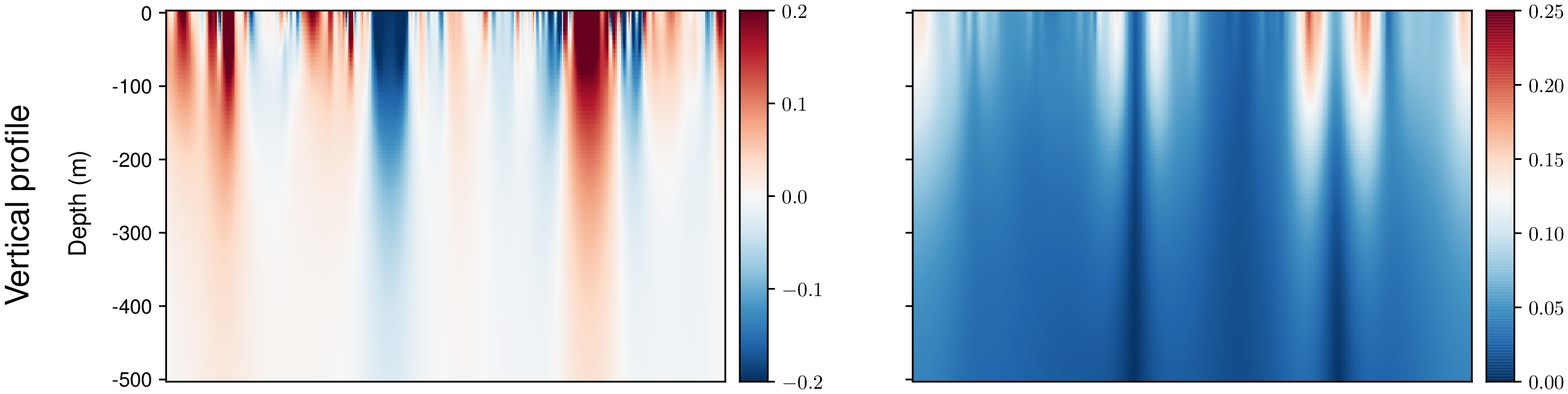}}
\caption{Flow initial condition for the storm experiment. Left: vertical relative vorticity normalized by the Coriolis frequency. Right: magnitude of the geostrophic velocity. Up: surface view. Down:  $xz$ section of the top 500 meters  taken along the horizontal line drawn in the top panels.} \label{IC}
\end{figure*}

The turbulent mesoscale field grows from the quasi-geostrophic baroclinic instability of the base flow $U(z)$. This growth is eventually halted by the bottom friction produced by a 60m-deep Ekman bottom layer. The spin-up  ends once the  eddy energy reaches statistical stationarity. At this point, we remove the energy-injecting base-state to better isolate wave-eddy interactions.

\begin{figure*}[h]
\centering
\includegraphics[trim=0 0 0 0, clip, width=.499\textwidth]{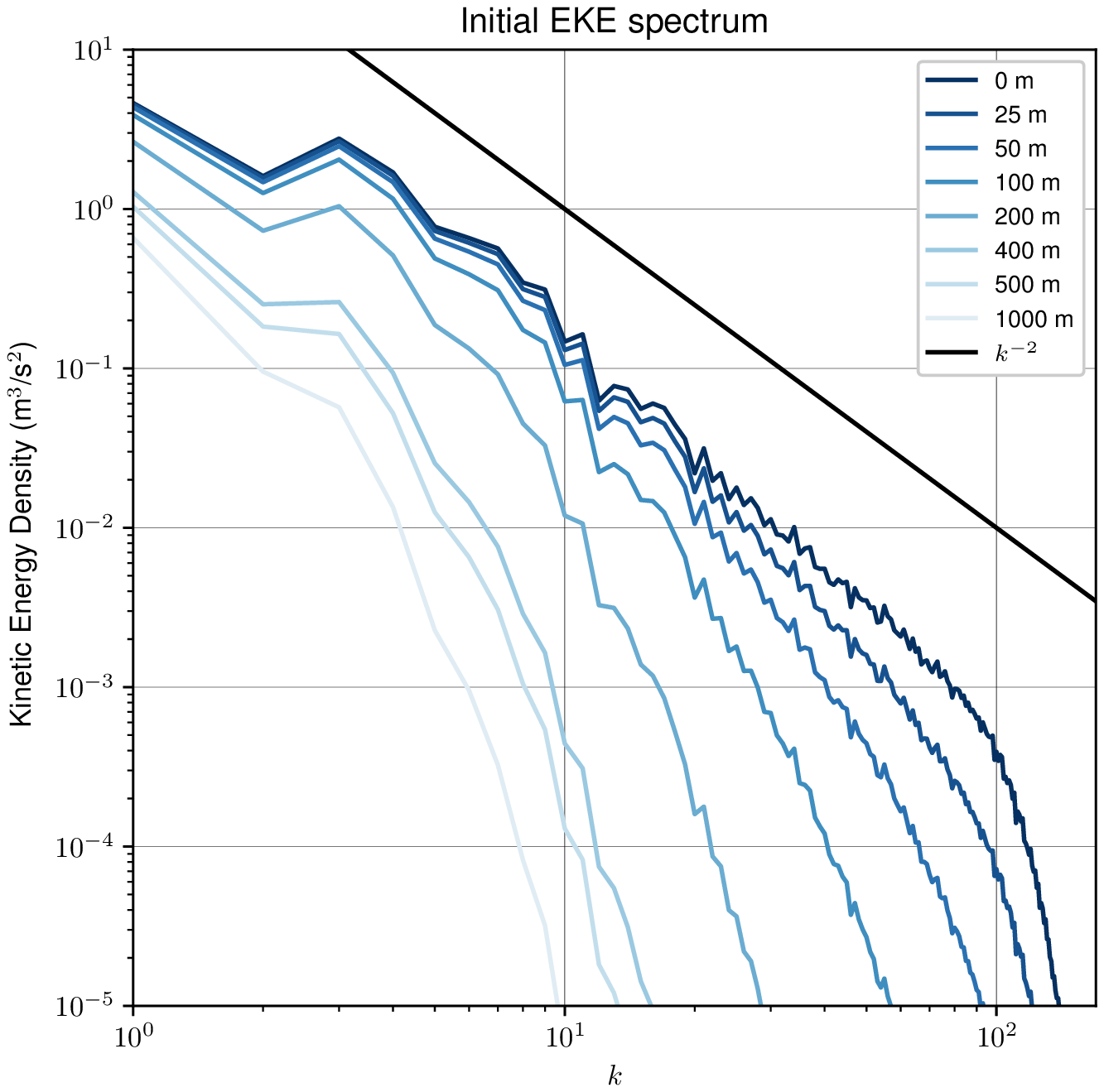}
\includegraphics[trim=0 0 0 0, clip, width=.492\textwidth]{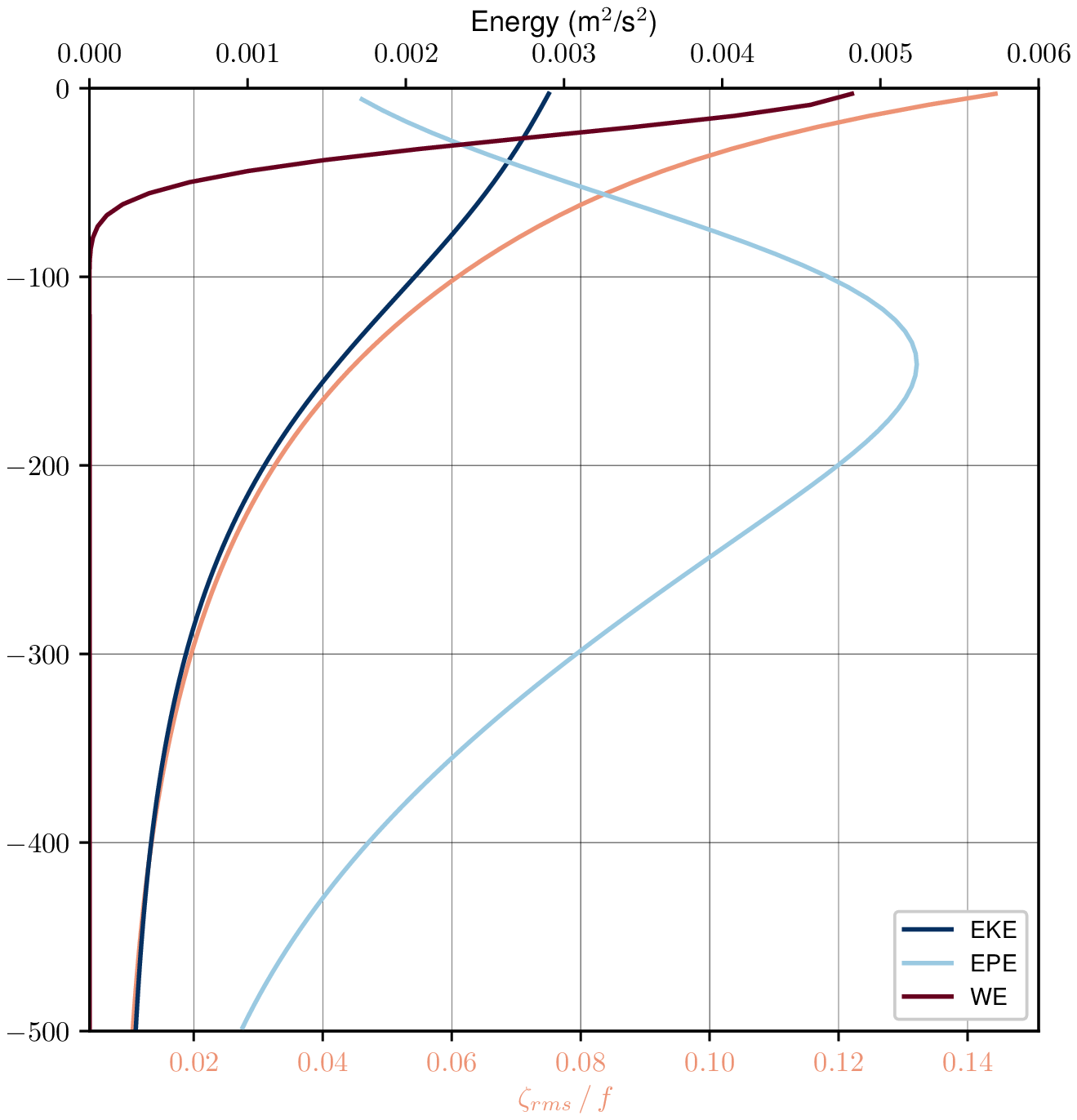}
\caption{Initial condition for the simulations. Left: horizontal-wavenumber eddy kinetic energy spectrum at various depths. Right: vertical profile of eddy kinetic (EKE) and potential (EPE) energy,  wave energy (WE, using $u_0=10$ cm s$^{-1}$) and normalized root-mean-square (rms) vorticity.}
        \label{IC2}
\end{figure*}

Figure \ref{IC} displays the snapshots of the equilibrated mesoscale field used as an initial condition for the storm experiments. The surface flow is a realization of surface quasi-geostrophic turbulence \citep{johnson,blumen,lapeyresqg}. Figure \ref{IC} shows secondary roll-up of filaments \citep{held1995}. Furthermore, the left panel of figure \ref{IC2} shows a shallow $k^{-2}$ kinetic energy spectrum at the sea-surface \citep{pierrehumbert1994} which rapidly steepens and weakens with increasing depth \citep{sb2013,onqg,onbq}. In the bottom panels of figure \ref{IC}, features with larger horizontal scales  have deeper vertical penetration scales. This is  consistent with surface quasi-geostrophic dynamics:  features with horizontal wavenumber $k$ decays exponentially with a vertical scale $f/N k $ \citep{ts2006}.

The right panel of figure \ref{IC2} shows the vertical profile of horizontally-averaged energy and vorticity for the initial condition. The vorticity and velocity fields decay vertically within the first few hundred meters (only the top 500m are shown, but the domain is 3km deep). The vorticity-based root mean square (rms) Rossby number, $Ro = \zeta_{\mathrm{rms}}/f$, reaches up to about 0.14 at the surface, which is comparable with the value of 0.10 in the simulations of \cite{danioux2008propagation}.  

\subsection{Phase II: Setting up a storm}

\cite{pollard_millard1970} found strong agreement between inertial currents from moored observations and those calculated from a simple slab model of the mixed layer. In this slab model, the mixed layer behaves as a harmonic oscillator with a resonant frequency $f$ (neglecting the artificial damping term). Mid-latitude storms typically have a strong inertial component (D'Asaro 1985) and thus cause the mixed-layer to resonate at its natural frequency. 

In our idealized scenario we assume that a passing storm  impulsively  excites an inertial oscillation throughout the mixed layer without influencing the geostrophic flow. Such events have been reported during the Oceans Storms Experiment --- see, for instance, the second storm analyzed in \cite{dohandavis}. Atmospheric storms have horizontal scales on the order of a thousand kilometers, so we shall assume that the wind-generated inertial oscillation is horizontally uniform in our 222 km $\times$ 222 km domain (about 2 degrees of longitude squared). As a result, this idealized scenario can be translated into an initial value problem for the wave field, with  an initial condition that is  a horizontally-uniform, mixed-layer-confined unidirectional current:
\begin{gather}
u =  u_0 \, \text{e}^{-z^2/\sigma_w^2} , \quad v=w=b=0, \label{waveIC}
\end{gather}
where $u,v,w$ and $b$ are the wave velocities and buoyancy. The parameter $\sigma_w$ is a proxy to the mixed layer depth. The initial wave field is given by \eqref{waveIC} with $\sigma_w = 50$ m: see  the red curve of the right panel of figure \ref{IC2}. The storm strength is mediated through the surface velocity, $u_0$. Unless otherwise specified, we use the typical value $u_0 = 10$ cm s$^{-1}$. The impact of storm strength is covered in section \ref{sec:feed}, where we consider values ranging from $u_0=0$ (equivalent to no feedback) to 40 cm s$^{-1}$.

\section{The QG-NIW model} \label{sec:model}

The previous section described  the initial condition used for eddy and wave fields; here we cover their subsequent evolution. In essence, the QG-NIW model couples the \ybjp{} equation for near-inertial waves \citep{ybjp} with the traditional quasi-geostrophic (QG) equation:
\begin{gather}
\p_t q +  J(\psi,q) = \mathcal{D}_q,  \label{qteq} \\
\p_t\LLp A + J(\psi,\LLp A) +  \halfi \lap \psi \, \LLp A+ \tfrac{\ii f}{2} \lap A   =  \mathcal{D}_{\LLp A}, \label{ybjp}
\end{gather}
where $i = \sqrt{-1}$ is the imaginary unit, $J(a,b) = a_x b_y - a_y b_x$ is the horizontal Jacobian, $\lap = \partial_x^2 +  \partial_y^2$ is the horizontal Laplacian. In \eqref{ybjp} we have introduced two frequently-occurring operators,
\beq
\LL \defn \frac{\partial}{\partial z} \left( \frac{f^2}{N^2} \frac{\partial}{\partial z} \right), \qquad \text{and} \qquad \LLp \defn  \LL +  \frac{1}{4} \lap. \label{ls}
\eeq
The complex field $A$ relates to the backrotated velocity of the near-inertial wave field --- see \eqref{la}  below --- while $\psi$ and $q$ are the streamfunction and potential vorticity of the balanced flow and $\mathcal{D}$ represents small-scale dissipative processes. 

\subsection{Wave evolution}

Wave evolution is dictated by  the \ybjp{} equation \eqref{ybjp}. Reading terms from left to right, \ybjp{} describes the evolution of the near-inertial wave envelope due to advection, refraction, dispersion and dissipation. In this framework the horizontal wave velocities, $u$ and $v$, are combined into a single complex field $A(x,y,z,t)$ with the fast  inertial rotation removed,
\beq
\LL A \defn (u + iv) \text{e}^{ift}. \label{la}
\eeq
Since waves are assumed near-inertial, the wave envelope \eqref{la} evolves slowly compared with the inertial period. In the original YBJ model \citep{ybj}, this wave envelope constitutes the prognostic variable from which all wave fields can be derived (like $q$ in the QG system). In \ybjp{} the prognostic variable employs the improved operator $\LLp$ instead of the original $\LL$ in \eqref{ls}. This tweak in the definition of the wave envelope makes  the physics more accurate and the numerics more docile whilst maintaining ease of implementation \citep{ybjp}. 

\subsection{Flow evolution}

The evolution of the balanced flow is dictated by \eqref{qteq}, which is identical  to the traditional quasi-geostrophic potential vorticity equation \citep{charney1948,salmonBook,vallis2017}. Here, however, $\psi$ and $q$ are defined as the streamfunction and potential vorticity of the \emph{Lagrangian}-mean balanced flow \citep{xv}:
\begin{gather}
q =  \lap \psi +  \LL  \psi  + \underbrace{\frac{i}{2f}J(\LLp A^*,\LLp A) + \frac{1}{4f}\lap |\LLp A|^2}_{q^w}, \label{qeq}
\end{gather}
where $A^*$ is the complex conjugate of $A$. The first two terms on the right-hand side of \eqref{qeq} are  the usual relative vorticity and stretching terms. The other term, $q^w$, is the wave feedback onto the flow \citep{xv,wagner2015}. Compared with the regular QG model, the inversion of $q$ for $\psi$ is trivially modified by subtracting $q^w$ from $q$. 

\cite{xv} showed that with strong waves it is the Lagrangian-mean flow, not the  Eulerian-mean flow, that is in geostrophic balance. Thus $\psi$ in \eqref{qteq}, \eqref{ybjp} and  \eqref{qeq} is a streamfunction for the Lagrangian velocity and might be decorated with a superscript $\mathrm{L}$. To lighten the notation we have dropped this $\mathrm{L}$.

The difference between the Lagrangian-mean and Eulerian-mean flow variables --- the Stokes correction --- scales like $Ro\,$WE/EKE. The magnitude of this Stokes correction can be estimated by inspecting figure \ref{IC2} in which $\zeta_{rms}/f$ is a Rossby number. With the standard value  $u_0 = 10$ cm s$^{-1}$, we obtain a Stokes correction on the order of 20\%. This indicates that the Eulerian-mean and Lagrangian-mean definitions do not differ qualitatively: flow variables may be interpreted as typical Eulerian-mean quantities for our standard runs (all of section \ref{sec:prop}). For storms strengths of 20 cm s$^{-1}$ or more, however, the Stokes ``correction" may be as large as its Eulerian-mean and Lagrangian-mean counterparts. We return briefly to these strong storms in section \ref{sec:feed}.

Finally, note that \eqref{qteq} is devoid of forcing terms. The end of the spin-up phase corresponds to eddy energy reaching statistical equilibrium in absence of waves  (\emph{e.g.} figure \ref{IC}). At this point we remove the Eady terms: base-state velocity, meridional gradients, and Ekman friction are set to zero. The elimination of these sources and sinks of eddy energy isolates the potential effect of near-inertial waves on eddies (section \ref{sec:feed}).

\subsection{Energetics}

In the \ybjp{} model, wave action and total energy are equivalent  \citep{ybjp}:
\beq
\text{WE} \defn \tfrac{1}{2} |\LLp A|^2. \label{WEdef}\\
\eeq
Upon multiplying the inviscid version of \ybjp{} equation \eqref{ybjp} with $\LLp A^*$, adding the complex conjugate, and performing a volume average, we find: 
\beq
\frac{d}{dt} \laa \text{WE} \raa = 0,  \label{we_cons} 
\eeq
where brackets $\laa \raa$ denote volume averaging. In the absence of dissipation and forcing, total wave energy is conserved.

\subsection{Numerical details}

The QG and \ybjp{} equations \eqref{qteq}-\eqref{ybjp} are solved using the same numerical methods. Both are pseudo-spectral in the $x$ and $y$ directions, allowing horizontal derivatives to be computed with spectral accuracy. The 2/3 rule is used to remove aliased modes. Vertical derivatives are approximated with second-order centered finite differences. The resolution used is $512^3$. Since the domain is $222 \times 222$ km in the horizontal and 3 km deep, this gives $\Delta x \approx 433$ m and $\Delta z \approx 6$ m, uniformly spaced. Time integration is accomplished with the leap-frog scheme with weak time diffusion \citep{asselin1972}. Dissipation occurs through a mixture of horizontal diffusion and hyperdiffusion:
\begin{align}
\mathcal{D}_{q} & = \nu_1 \lap^6 q + \nu_2 \lap^2 q, \\
\mathcal{D}_{\LLp A}  &= \nu_1 \lap^6 \LLp A,
\end{align}
with $\nu_1 = 3.2 \times 10^{25}$ m$^{12}$s$^{-1}$ and $\nu_2 = 9.5 \times 10^3$ m$^4$s$^{-1}$.

\section{Wave propagation} \label{sec:prop}

We begin by looking at how eddies distort the wave field and allow penetration into the ocean interior. We found that the main storyline is not qualitatively affected by the storm strength, nor even by wave feedback via $q^w$ in \eqref{qeq}. This section thus focusses on the standard case with wave feedback and a wind storm of $u_0=10$ cm/s. We postpone our discussion of storm strength dependence and wave feedback onto eddies to section \ref{sec:feed}.

\subsection{Early refractive phase} \label{sec:ref_phase}

\begin{figure*}
\centering
\includegraphics[trim=30 140 40 120, clip, width=.7\textwidth]{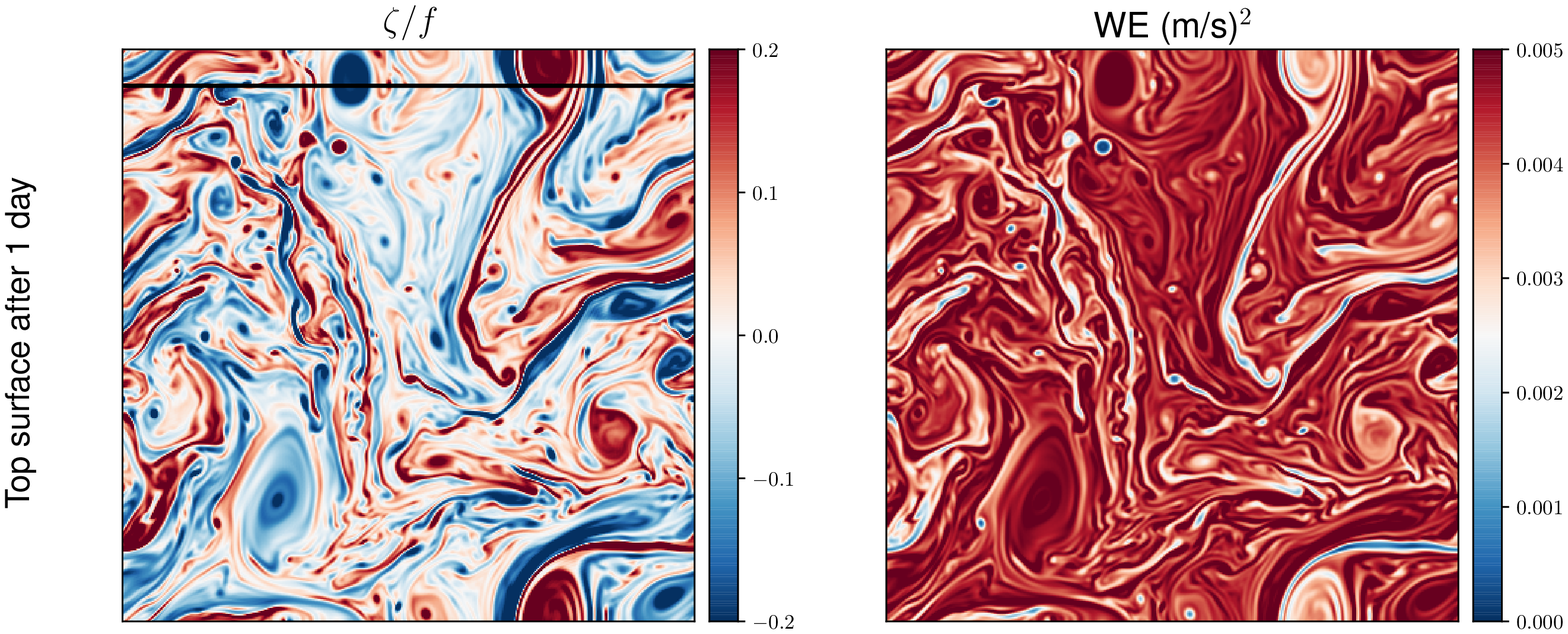}
\includegraphics[trim=30 140 40 140, clip, width=.7\textwidth]{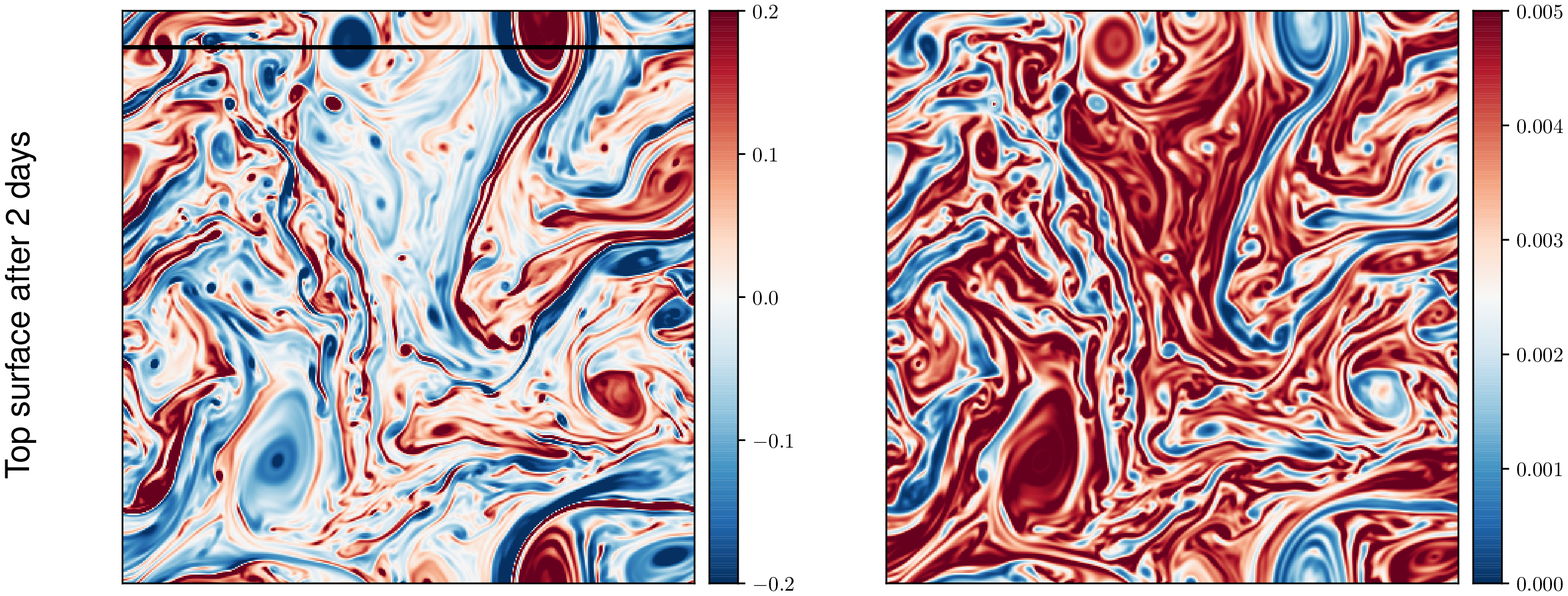}
\includegraphics[trim=30 140 40 140, clip, width=.7\textwidth]{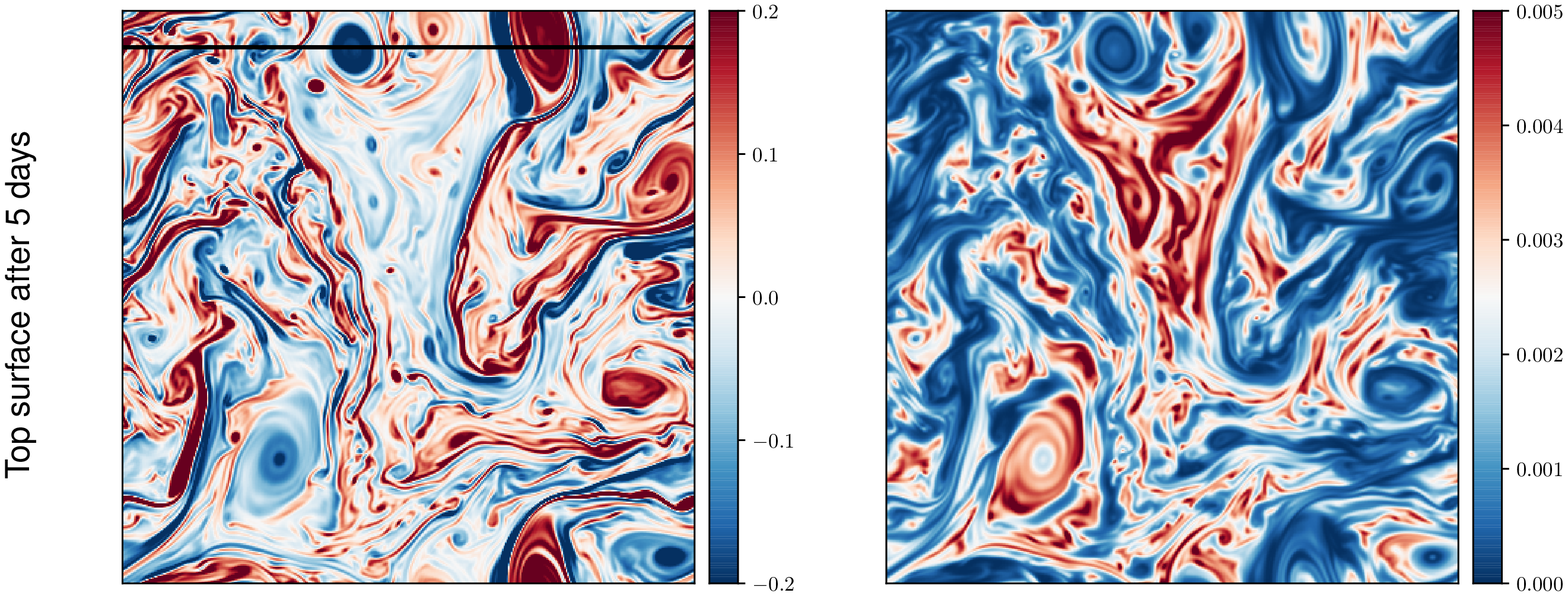}
\includegraphics[trim=30 140 40 140, clip, width=.7\textwidth]{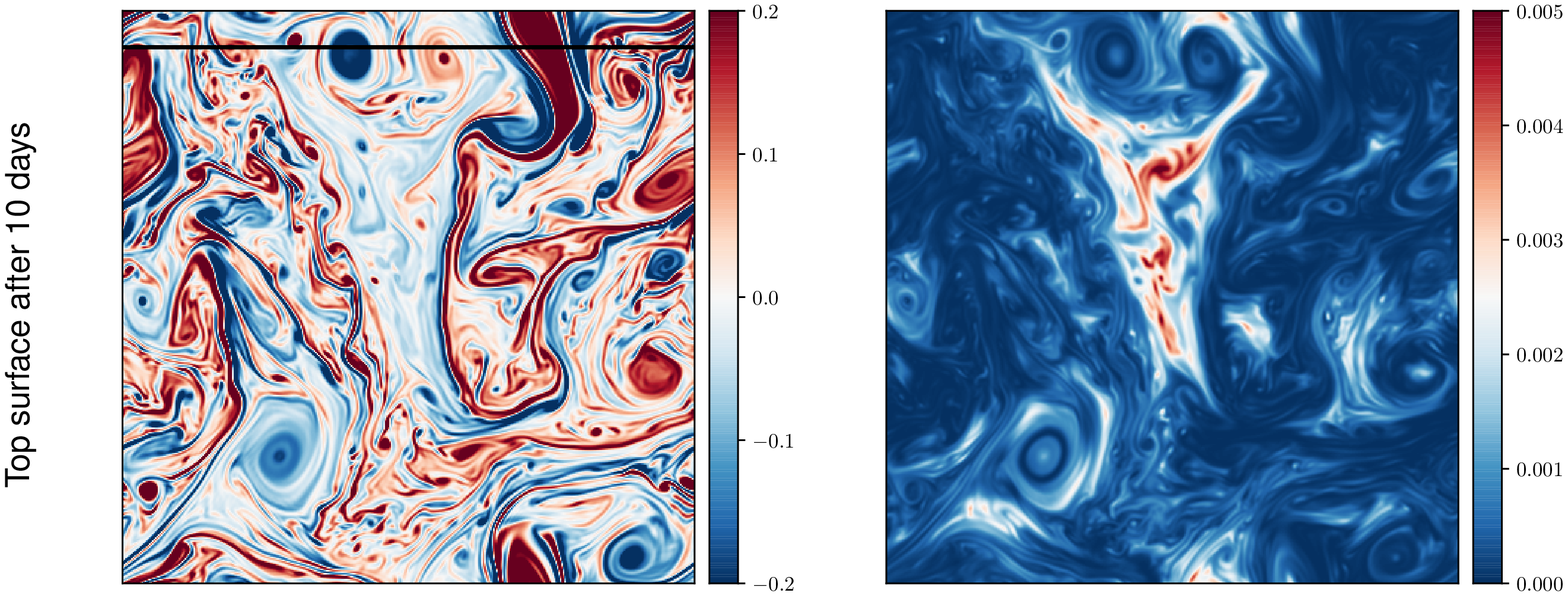}
\includegraphics[trim=30 140 40 140, clip, width=.7\textwidth]{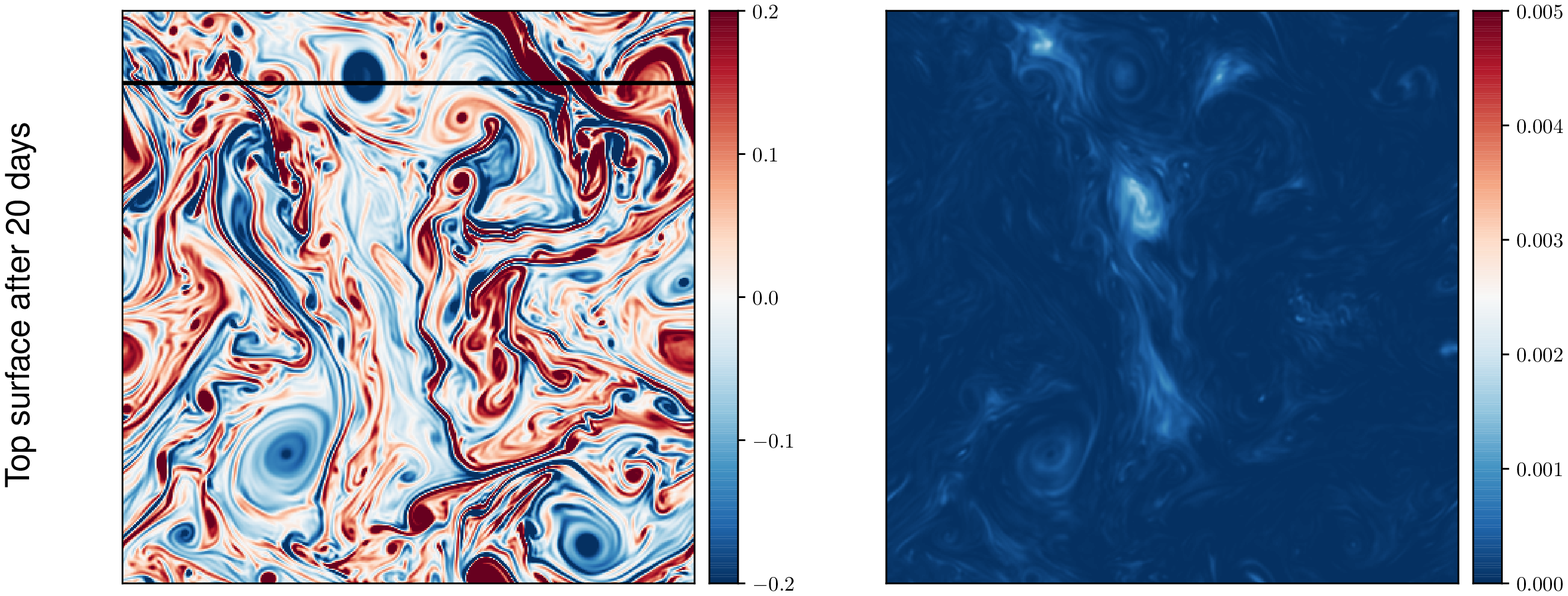}
\caption{Time series of vorticity and total wave energy density at the surface. Horizontal lines indicate the location of vertical slices for figure \ref{wke_zeta_ver_shallow}.}
 \label{wke_zeta_long}
\end{figure*}

Figure \ref{wke_zeta_long} shows snapshots of surface wave energy (WE) and normalized vertical vorticity, $\zeta/f$, after 1, 2, 5, 10 and 20 days (from top to bottom). The eye is immediately arrested by the rapid imprinting of the vorticity scales onto the wave field, leading to a collapse of the near-inertial wave horizontal scale.

This early refractive phase has been studied  by \cite{klein2004}. For an initially pure inertial wave (horizontally-uniform amplitude and phase), the small-time balance in \eqref{ybjp} is 
\beq
\partial_t \LL A \approx -\tfrac{\ii}{2} \zeta \LL A,
\eeq
where next-order \ybjp{} corrections are omitted for convenience. The early-time wave solution is therefore:
\beq
\LL A \approx \LL A_0 \,  \text{e}^{-\tfrac{i}{2} \zeta t}, \label{earlytime}
\eeq
where the 0 subscript denotes the horizontally-uniform initial condition in \eqref{waveIC}. Let's decompose the complex wave envelope in terms of real-valued amplitude, $R$, and phase $\theta$: $\LL A = R \, \text{e}^{i\theta}$. Taking the absolute value of \eqref{earlytime} indicates that refraction leaves $R$ unchanged. Instead, at early times, refraction causes a shift in the wave phase: 
\beq
\theta \approx \theta_0 - \tfrac{1}{2} \zeta t \label{theta_shift}.
\eeq
Using altimetry and surface drifter data, \cite{elipot} found that phase shifts approximately follow $-0.39 \zeta t$, which is reasonably near to the refractive shift predicted here (and by \cite{kunze1985} and others).

To obtain the evolution of the wave scale, one takes the horizontal gradient of \eqref{theta_shift},
\beq
\mathbf{k} \defn \nabla \theta \approx - \tfrac{1}{2} \nabla \zeta \, t \label{k_shift}.
\eeq
Thus, refraction leads to the reduction of the wave horizontal scale: this is the essential ingredient that increases the vertical group velocity resulting in  propagation. Including dispersive effects, \cite{klein2004}  further showed that the early-time amplitude of the near-inertial wave distributes like the Laplacian of vorticity; see also \cite{kleintreguier}. Note however that $\lap \zeta$ is strongly anti-correlated with $\zeta$ \citep{elipot}, which explains the strong anticorrelation between wave energy and vorticity evident  at 1 day and 2 days in figure \ref{wke_zeta_long}.

This attraction of wave energy into  anticyclonic regions, and repulsion from cyclonic regions, is a well-known phenomenon, occurring both in realistic primitive-equation models \citep{danioux2008propagation} and observations \citep{elipot}. Explanations have been proposed relying on the broadening of the allowable frequency band in negative vorticity regions and subsequent trapping of rays \citep{kunze1985}, appeals to the quantum analogy between energy wells and negative vorticity \citep{balmforth1998}, and more recently, a  conservation law of the YBJ model that applies to  near-inertial waves in a  steady  flows \citep{DVB2015}.

\subsection{Later-time wave energy-vorticity correlation}

Surface wave energy is repelled from cyclones and attracted by anticyclones during the early-time refractive phase. At longer time, however, the story is more complicated. For instance, pick the negative (blue) and positive (red) vortices transected by  the black line in figure \ref{wke_zeta_long}. In the early stage (1 and 2 days) these vortices are associated with high (in the anticyclone) and low (in the cyclone) concentrations  of WE. But  at  5 days,  WE is gone from the negative vortex whereas the positive vortex has  drawn in some filaments of WE. Thus at 5 days, and at the sea surface,  there is more energy in the positive vortex than in the negative.  This is a result of vertical propagation: the WE initially focussed into the negative vortices has gone downwards (further discussion below). At 10 days, the small amount of  WE that remains at the sea surface is mostly in the region of weak vorticity.

\begin{figure*}[h!]
\centering
\includegraphics[trim = 40 0 0 0, width=.45\textwidth]{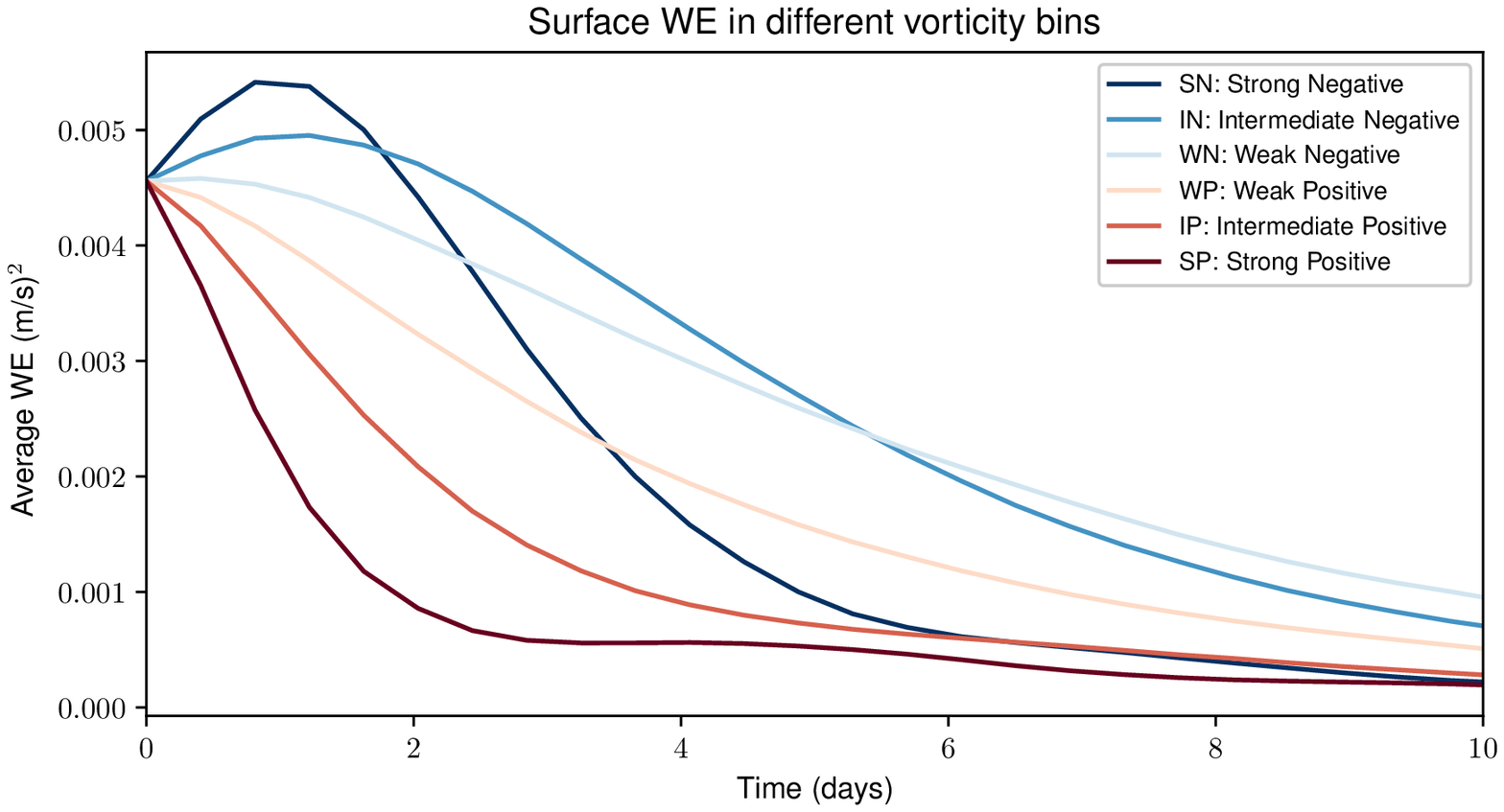}
\includegraphics[width=0.48\textwidth]{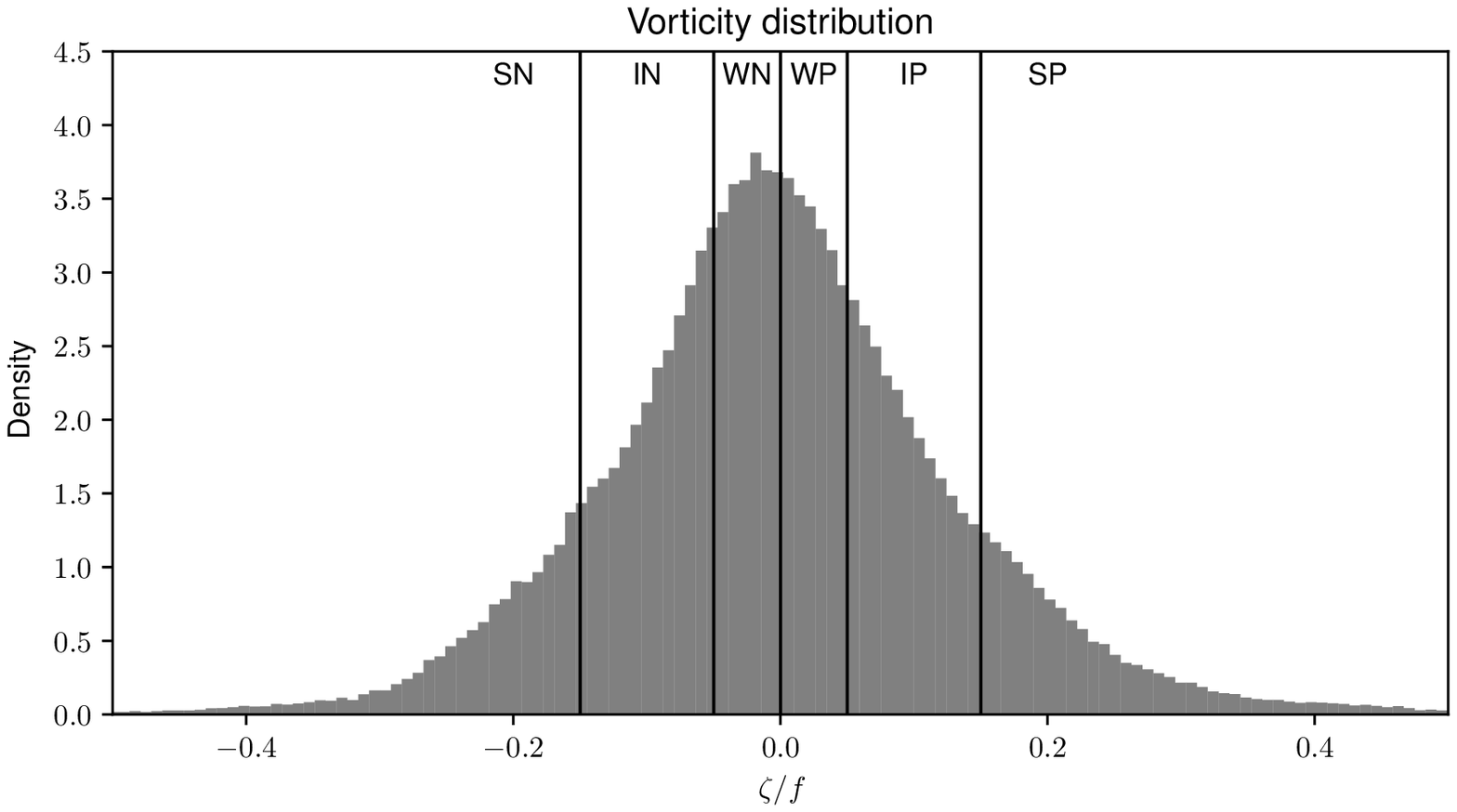}
\caption{Left: time series of the average of surface WE in regions of strong, intermediate and weak positive/negative vorticity. Right: initial distribution of vorticity (undergoes little change over time period shown). Boundaries between bins are defined as $\zeta/f = 0, \pm 0.05$,  and $\pm 0.015$.}
        \label{zeta_hist}
\end{figure*}

Figure \ref{zeta_hist} quantifies this evolution by displaying the average WE in different bins of surface vorticity, from strong negative to strong positive (see the right panel for the vorticity distribution and bin definitions). Early on, the more negative vorticity is, the more WE concentrates in this region. For the intermediate and strong positive regions there is a rapid decline of energy. After about 5 days, it is regions of weaker vorticity (and intermediate negative) that retain most of the sea-surface WE; see also the bottom panels of figure \ref{wke_zeta_long}. These weak-vorticity regions are also characterized by weak vorticity \emph{gradients} which, according to \eqref{k_shift}, are associated with a slower reduction of the horizontal scale and thus slower vertical propagation.

Using inverse excess bandwidth of near-inertial peaks in global drifter data,  \cite{elipot} estimated the average decay time of near-inertial waves at the surface as a function of the flow vorticity. Consistent with our simulations (in particular, figure \ref{wke_zeta_long}), they found that decay time is maximum for weaker vorticity, and it decreases faster with increasing positive vorticity than negative vorticity. 

\begin{figure*}
\centering
\includegraphics[width=.4\textwidth]{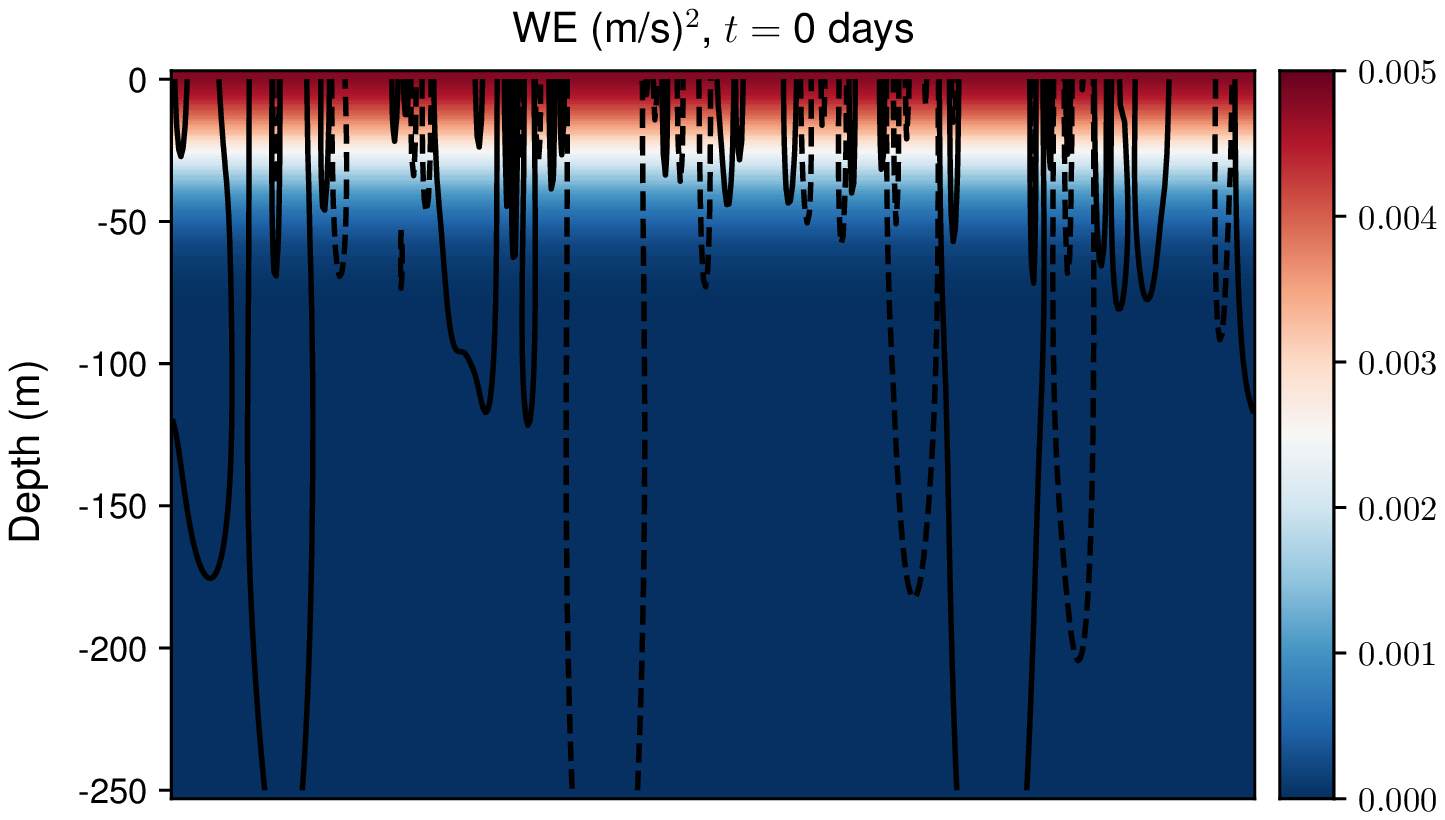}
\includegraphics[width=.4\textwidth]{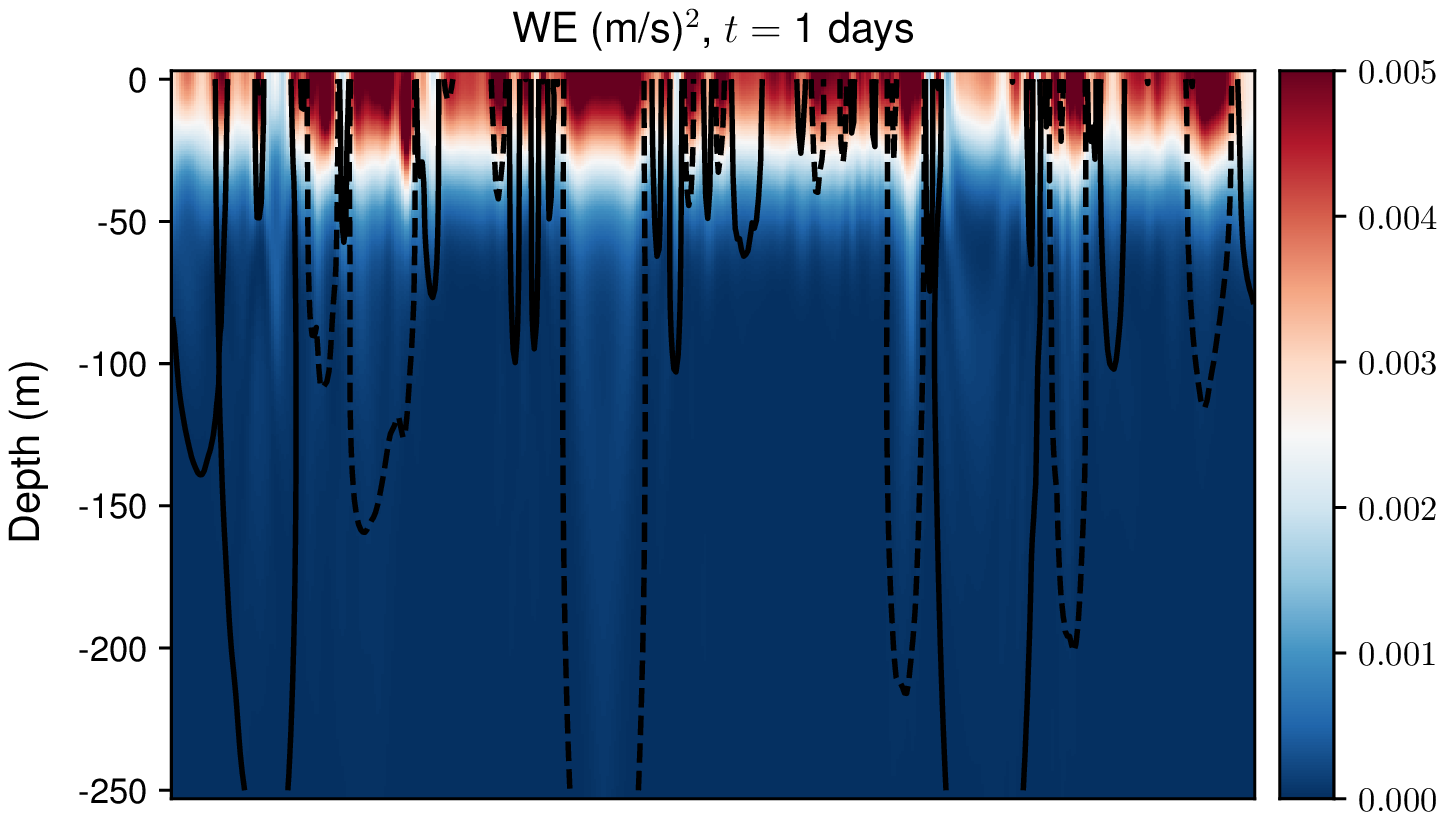}
\includegraphics[width=.4\textwidth]{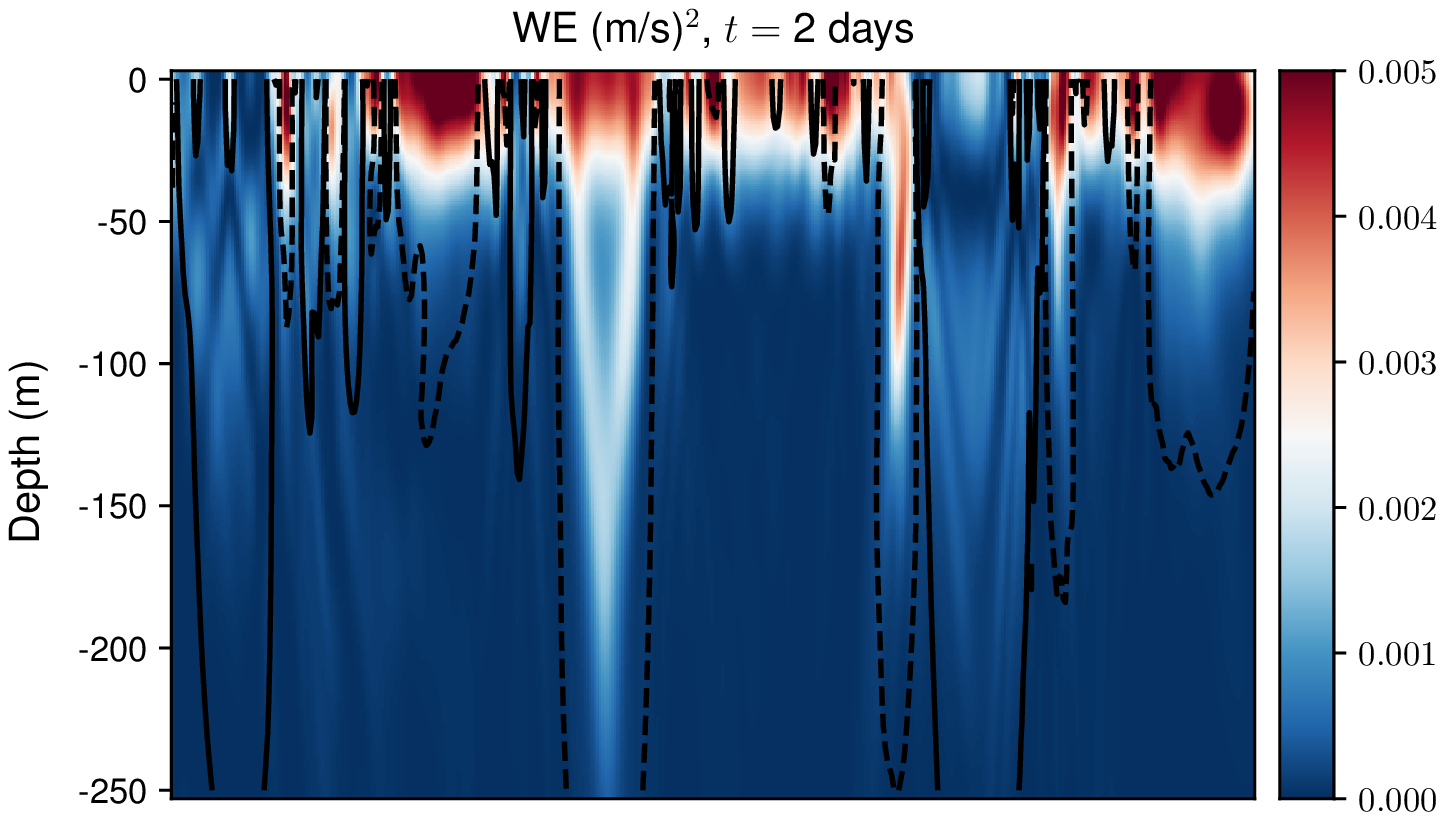}
\includegraphics[width=.4\textwidth]{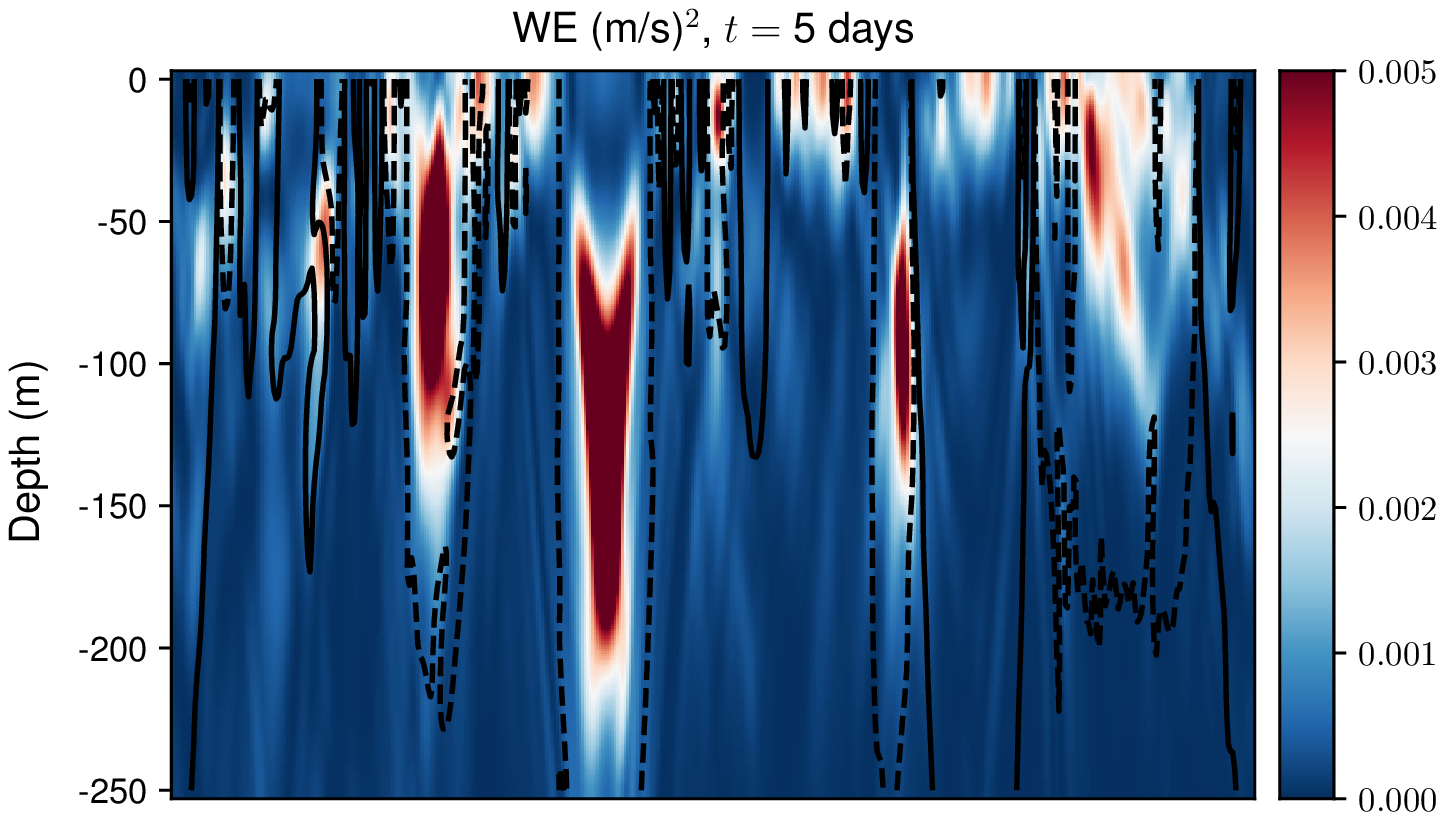}
\includegraphics[width=.4\textwidth]{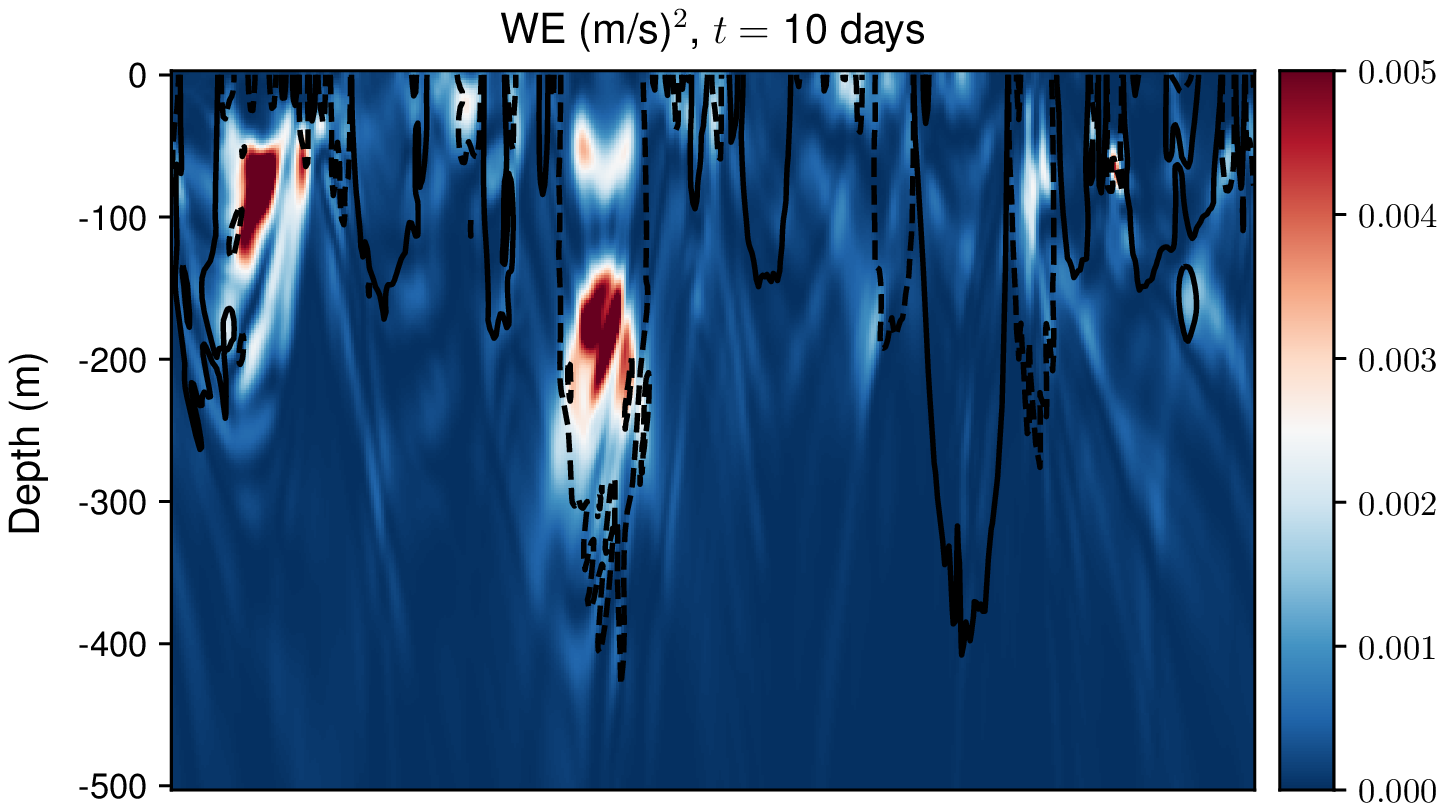}
\includegraphics[width=.4\textwidth]{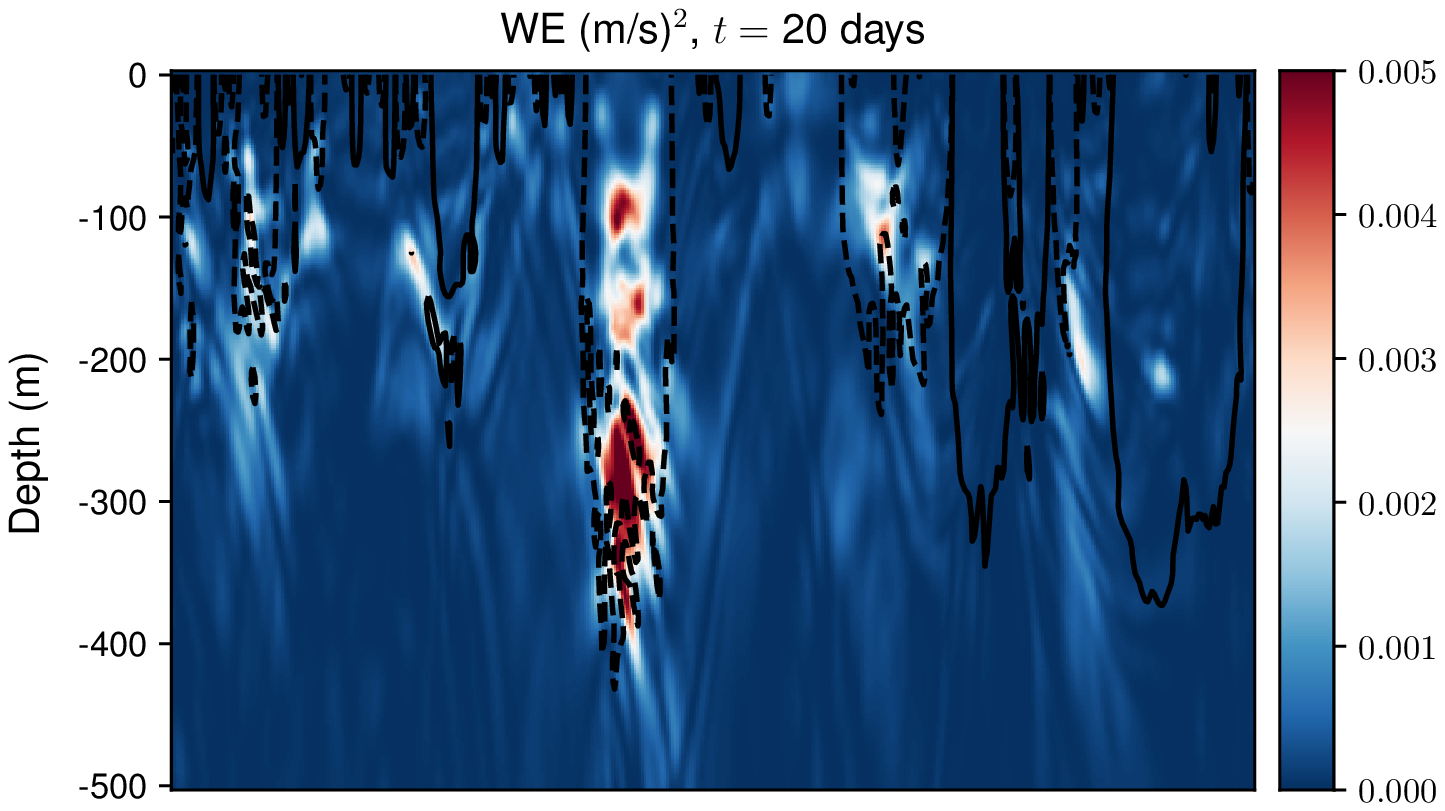}
\caption{$xz$ snapshots of WE with overlaid contours of vorticity (solid: $\zeta/f = 0.05$, dashed: $\zeta/f = -0.05$). The top 250 meters are shown for the 5 first days, then the top 500 meters are shown. The whole $x$ extent of the domain is shown. The $y$ location of slices is indicated by horizontal lines in the vorticity distributions of figure \ref{wke_zeta_long}.}
        \label{wke_zeta_ver_shallow}
\end{figure*}

\cite{elipot} also report that, on average, more surface wave energy populates anticyclonic than cyclonic regions. The difference, however, is small --- about a factor of two between the strongest positive and negative vorticity regions, with a relatively flat response for weaker vorticity (see their figure 15). This is consistent with a time-average of the $\zeta$-WE correlation  in figure \ref{zeta_hist}. Although the refractive phase initially causes a strong concentration of WE into anticyclones --- WE in strong  negative vortices is up to a factor of five larger than WE in strong positive vortices  --- the time-averaged surface correlation is rapidly diluted by the vertical propagation of waves into the ocean interior. We speculate that  in statistical steady state, in which near-inertial waves are intermittently forced by the passage of storms, the  distribution of WE will depend on the frequency of  storms that re-initiate the  early-time refractive phase. In general, we expect the time-averaged surface $\zeta$-WE correlation to be much less than predicted by $\zeta$-refraction alone.

\subsection{Inertial drainpipes} \label{sec:drain}

A main message of figure \ref{wke_zeta_long} is that WE vanishes rapidly from the sea-surface. Most of this loss at the surface is, of course,  because WE radiates into the ocean interior. Figure \ref{wke_zeta_ver_shallow} displays and $(x,z)$ section of WE at a fixed $y$ along the  black line overlaid on the vorticity plots in figure \ref{wke_zeta_long}; contours are overlaid to indicate regions of positive (solid) and negative (dashed) vorticity.

The picture is clear: WE leaves regions of positive vorticity within a day or two. This transfer is mostly lateral \citep{kunze1985,balmforth1998,leeniiler}. After 2 days there is significant downward propagation of energy guided along the cores of  anicyclonic (negative) vortices. Almost no subsurface WE is found in  cyclonic vortices. Anticyclones are wave guides that  drain WE downwards into the deeper ocean \citep{balmforth1998,leeniiler,zhai2005,danioux2008propagation}. After 10 or 20 days, WE  collects at the bottom of anticyclones. Using ray tracing, \cite{kunze1985} predicted this trapping of waves at the bottom of anticyclones as they encounter a critical layer --- a region defined by vanishing vertical group velocity as vorticity weakens. Wave trapping at the base of anticyclones  has frequently been observed \citep{kunze1984,kunze1986,oey2008stalling,joyce2013,martinez2019,kawaguchi2019}.

\begin{figure*}
\centering
\includegraphics[trim=30 140 30 130, clip, width=.7\textwidth]{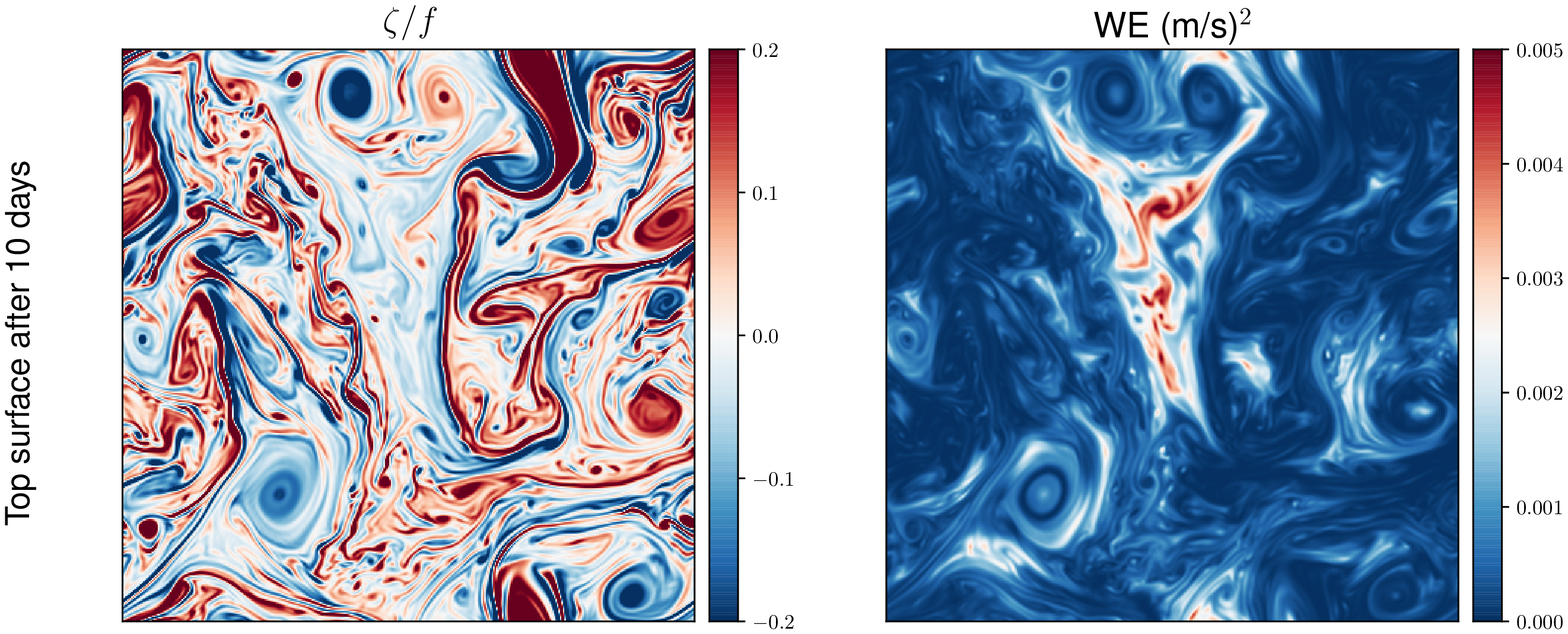}
\includegraphics[trim=30 140 30 140, clip, width=.7\textwidth]{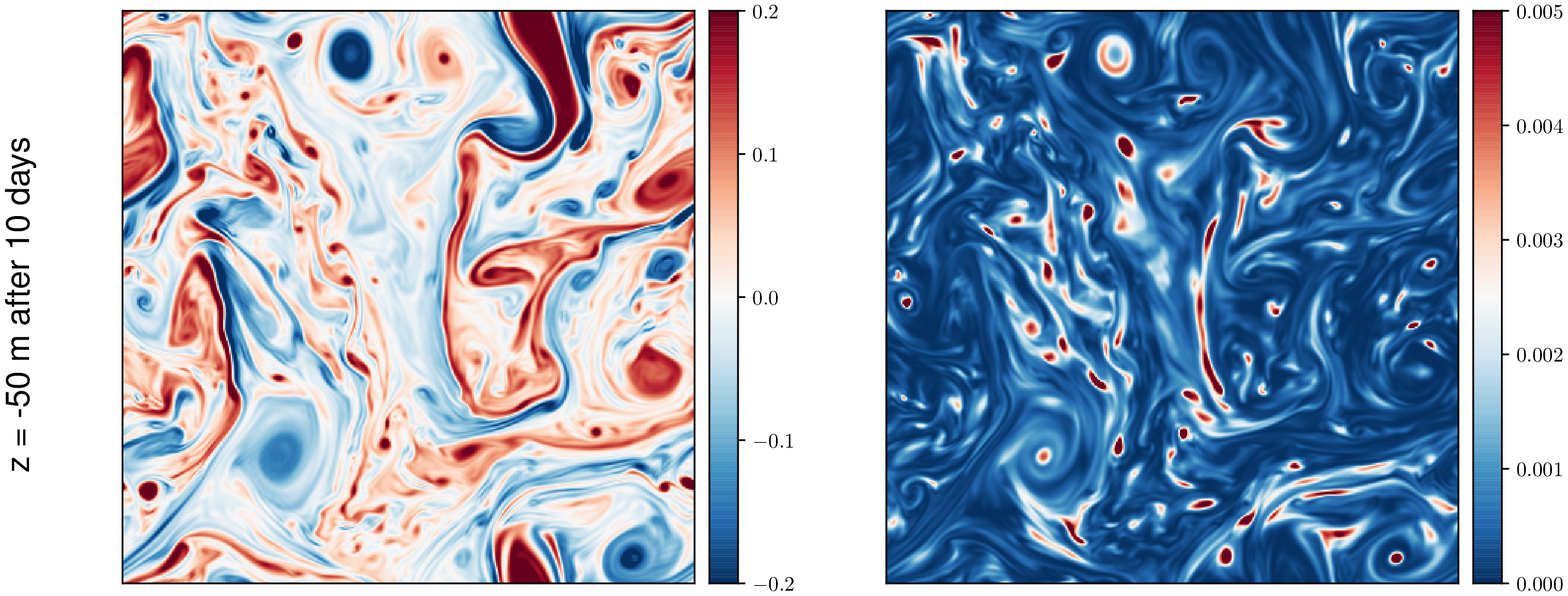}
\includegraphics[trim=30 140 30 140, clip, width=.7\textwidth]{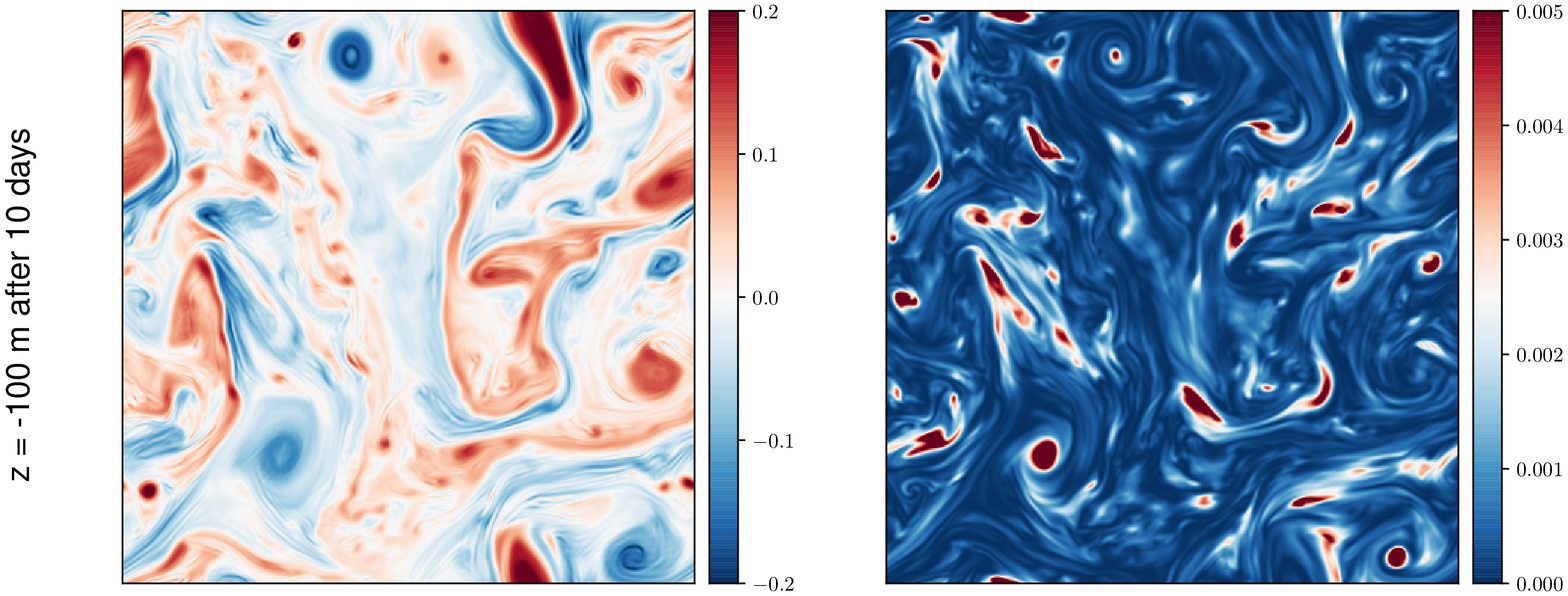}
\includegraphics[trim=30 140 30 140, clip, width=.7\textwidth]{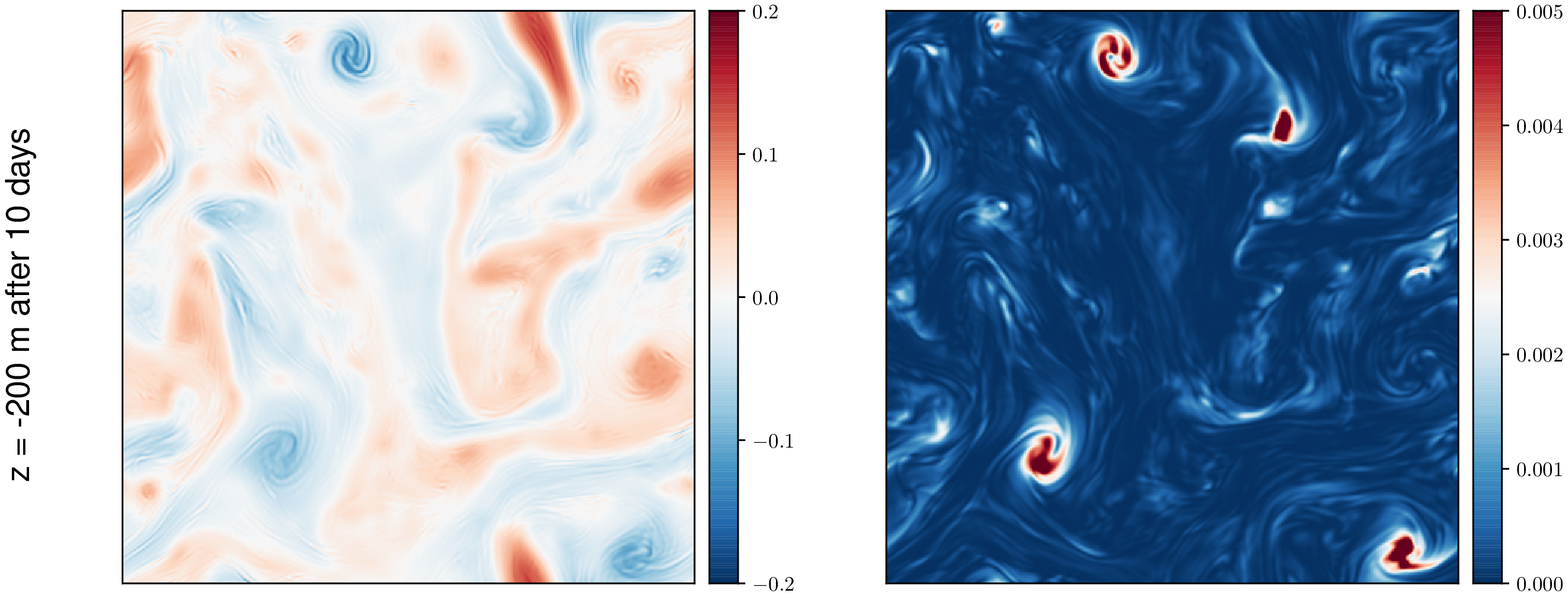}
\caption{Vorticity and WE after 10 days at various depths.}
        \label{wke_zeta_long10d}
\end{figure*}

Figure \ref{wke_zeta_long10d} shows horizontal cuts of $\zeta$ and WE at depths of 50, 100, 200 meters after 10 days. As seen in figure \ref{IC}, vorticity features with larger horizontal scales penetrate to greater depths. Like $\zeta$, WE is seen in larger-scale features at greater depths. Unlike $\zeta$, however, large-scale WE features are not seen at shallow depths: this is because WE collects at the bottom of anticyclones. For instance, at 50 m WE is only evident  in  the smallest vortices and filaments, at 100 m WE is found only in  intermediate-scale features and at 200 m WE only occupies  large anticyclonic cores. These observations provide further evidence that waves get trapped at the bottom of vortices as these weaken and form critical layers.

\subsection{Mixing}

As near-inertial waves propagate into the ocean interior, their vertical shear may come to exceed the stabilization provided by background stratification. In this case waves may break through shear instabilities and cause mixing. However our wave model filters out these instabilities \citep{ybj}. The necessary condition for  instability may nevertheless be diagnosed from  model output.

\begin{figure*}
\centering
\includegraphics[width=.4\textwidth]{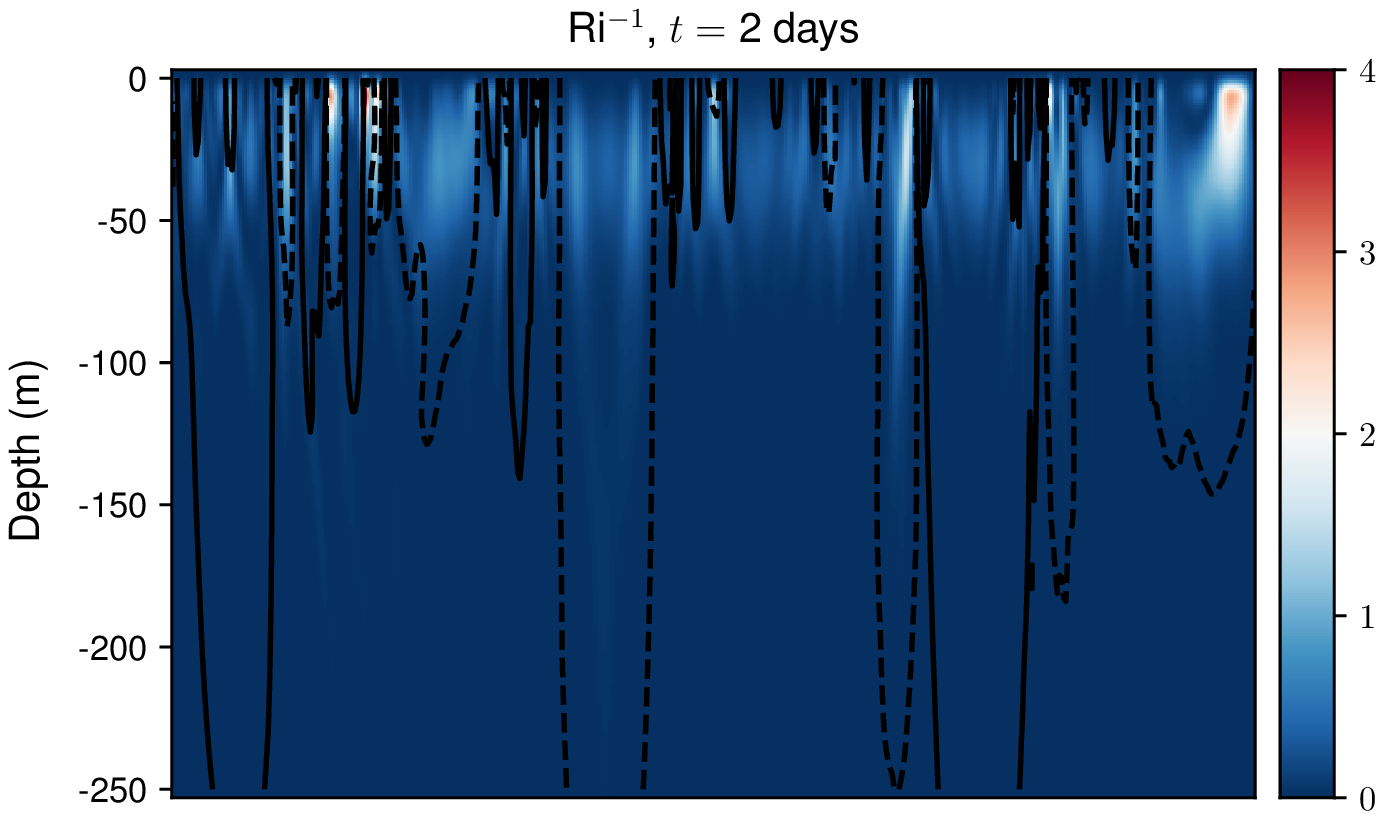}
\includegraphics[width=.4\textwidth]{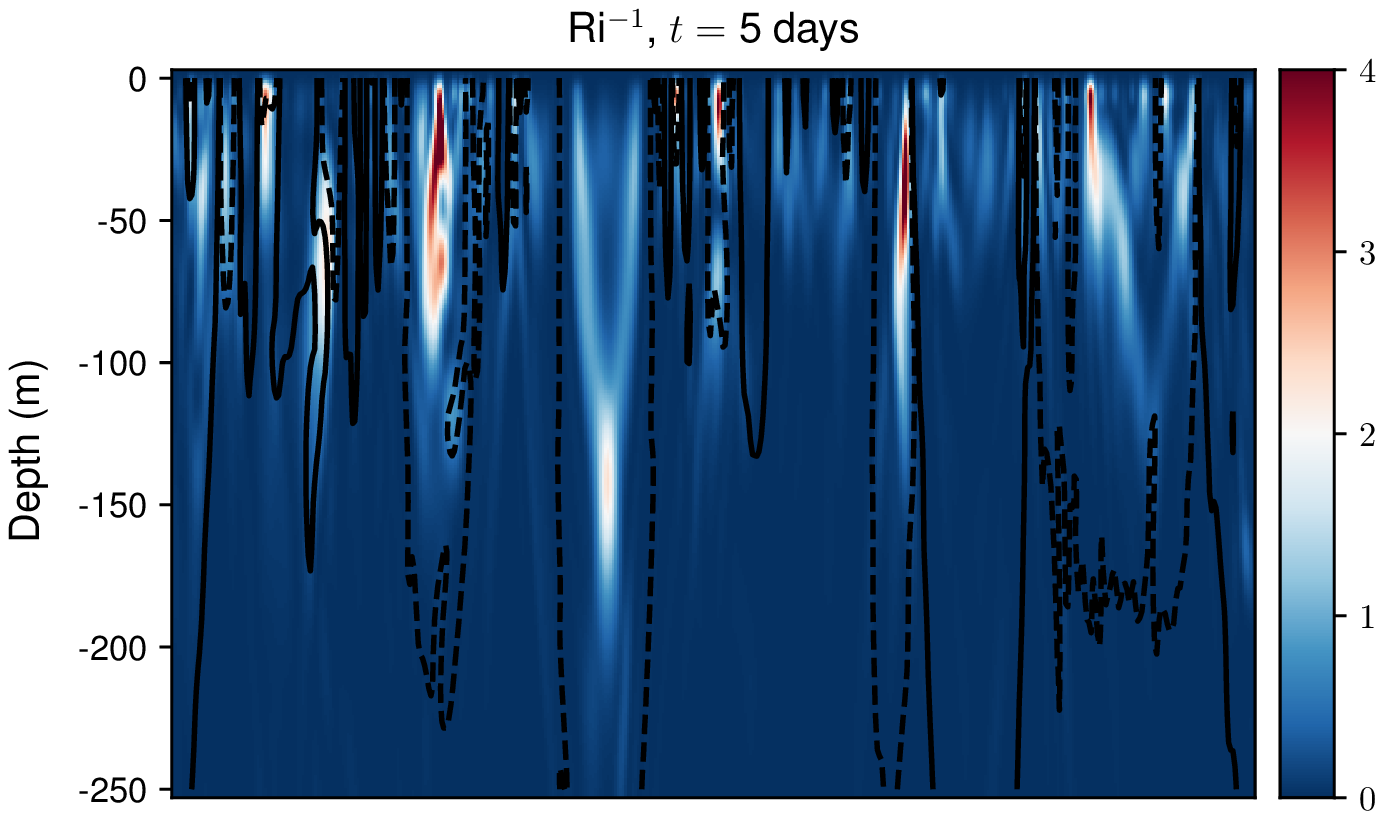}
\includegraphics[width=.4\textwidth]{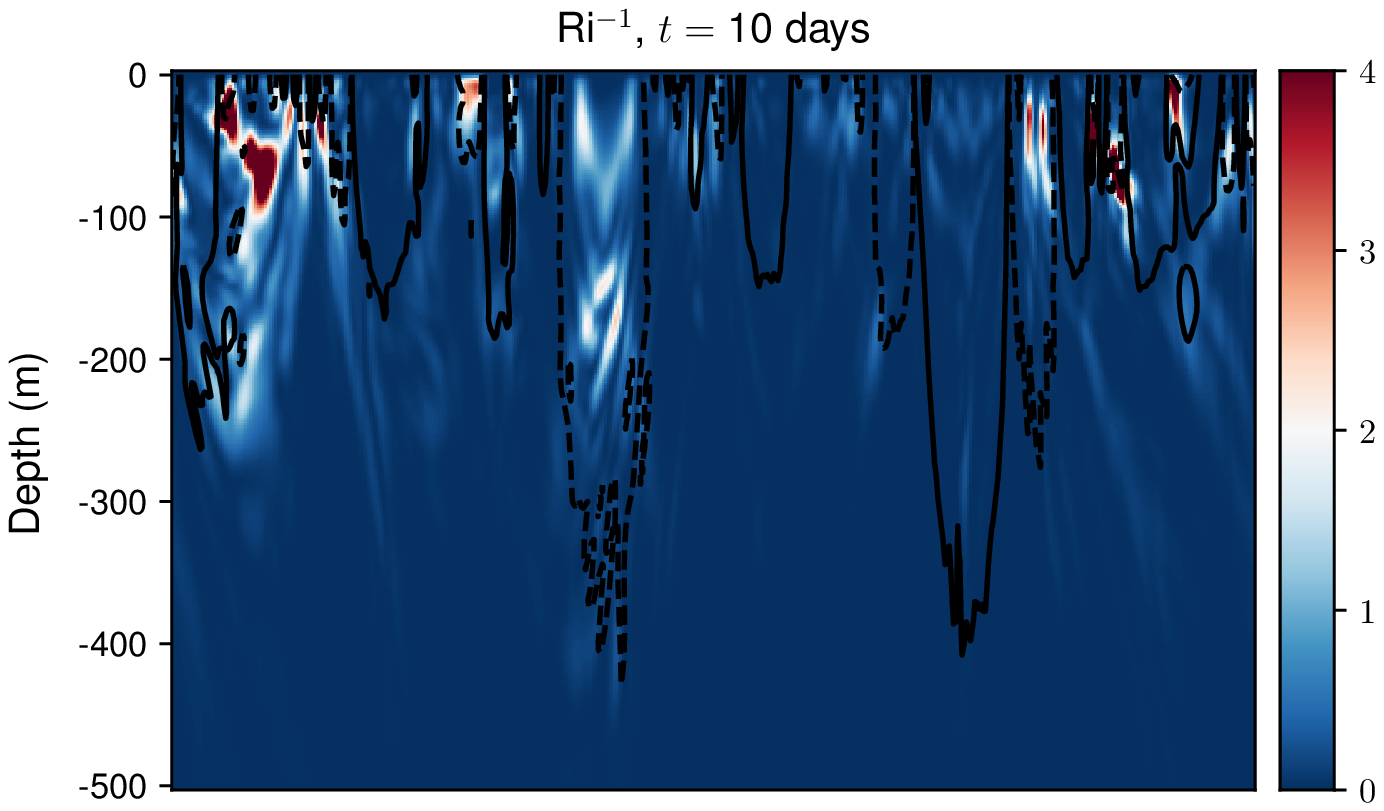}
\includegraphics[width=.4\textwidth]{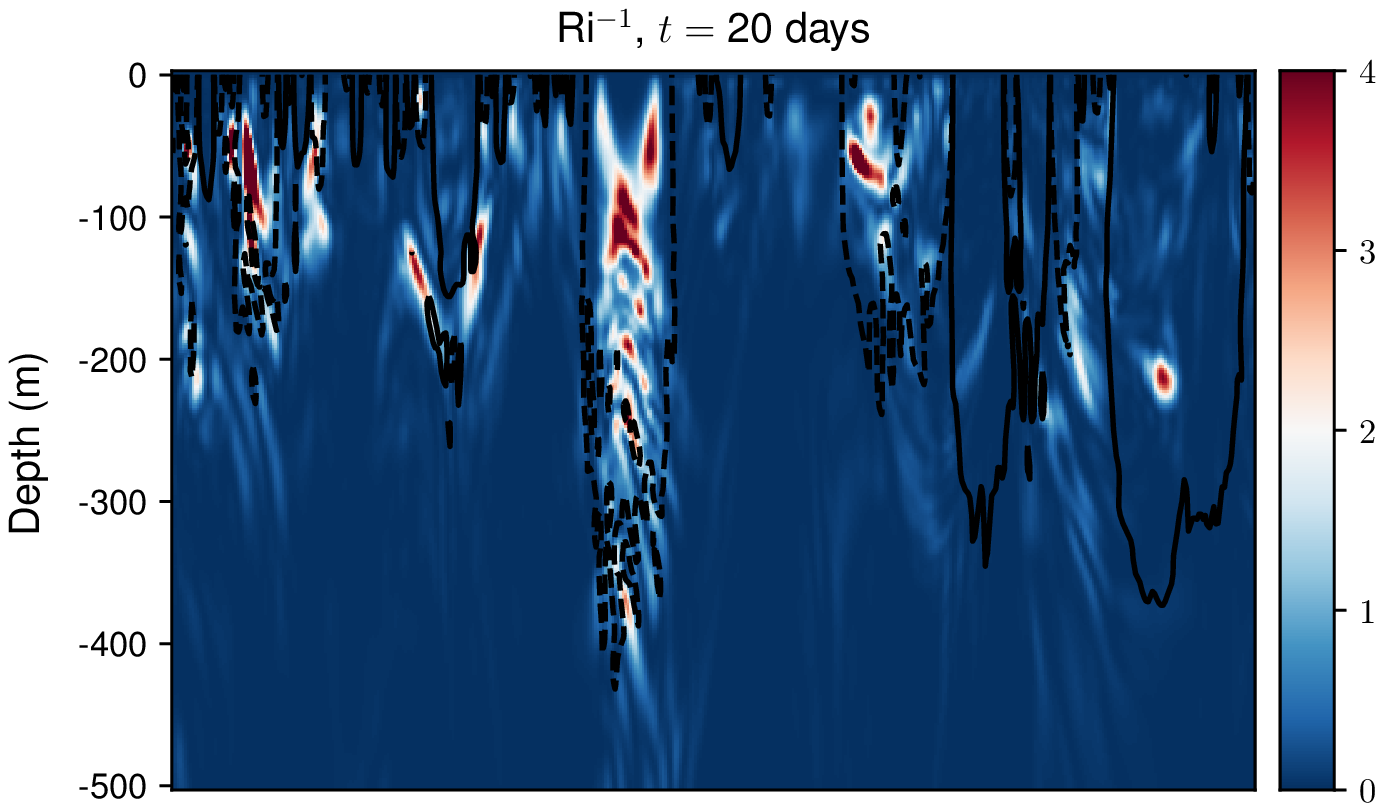}
\caption{$xz$ snapshots of the inverse Richardson number with overlaid contours of vorticity (solid: $\zeta/f = 0.05$, dashed: $\zeta/f = -0.05$). Values over 4 indicate potential shear instability. The top 250 meters are shown for the 5 first days, then the top 500 meters are shown.}
        \label{irn}
\end{figure*}

Figure \ref{irn} shows vertical cuts of the inverse Richardson number, 
\beq
Ri^{-1} \defn \frac{1}{2} \frac{u_z^2 + v_z^2}{N^2},
\eeq
which quantifies the relative stabilizing and destabilizing contributions of stratification and vertical shear. Values of 4 or more (saturated red) indicate that shear instabilities could occur, and thus provides a proxy to the presence of mixing. 

During first 10 days, mixing is mostly confined to the top 100 m, in spite of significant wave energy present at lower depths (compare with figure \ref{wke_zeta_ver_shallow}). This is because stratification is weaker near the surface (figure \ref{N2}). After 20 days, mixing extends throughout the whole depth of the large negative vortex.  Overall, mixing is collocated with wave energy --- in anticyclones --- echoing the observations of \cite{lueck1986dissipation} and \cite{kunze1995}.

\subsection{Depth distribution of wave energy}

\begin{figure*}
\centering
\includegraphics[width=19pc]{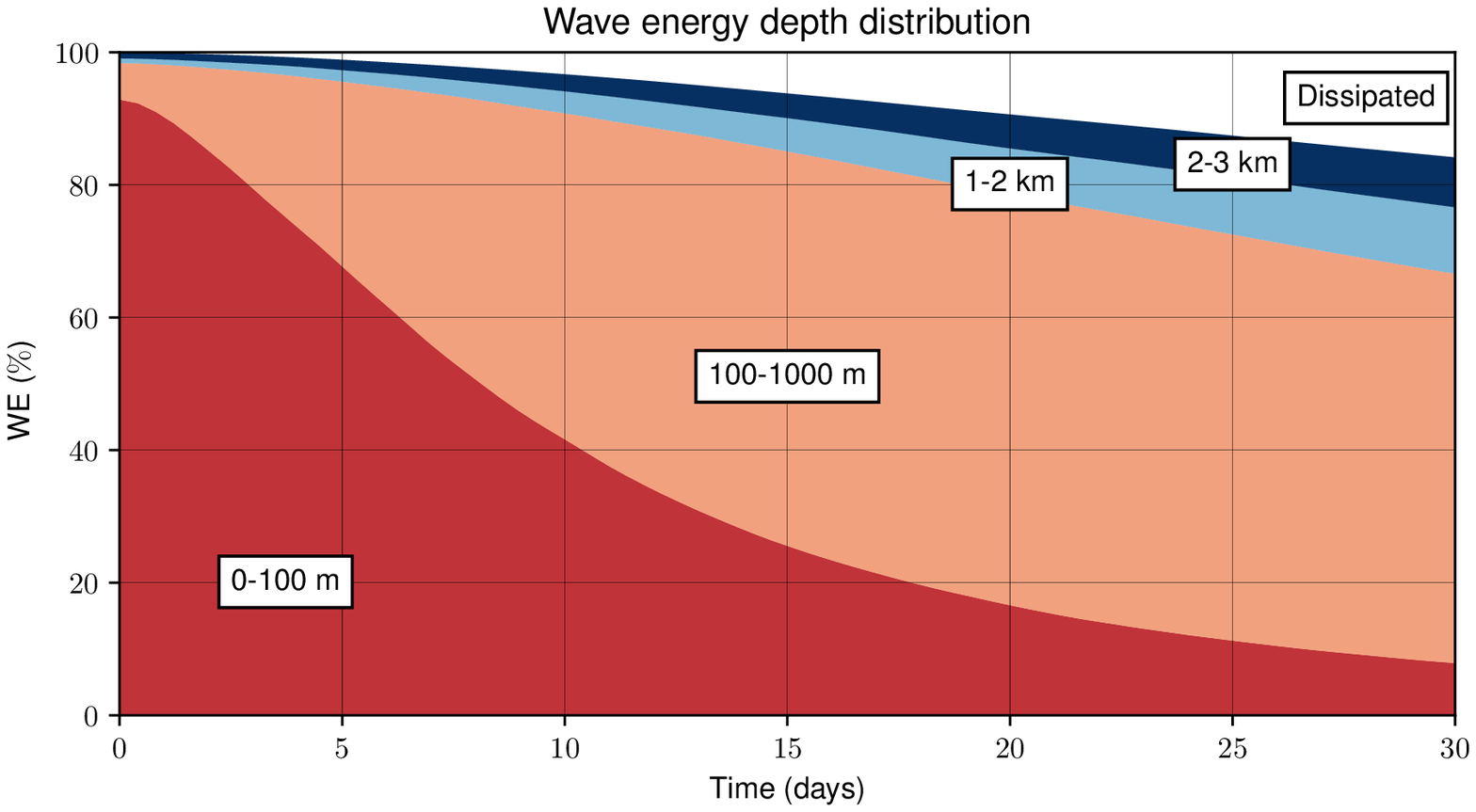}
\includegraphics[width=18.1pc]{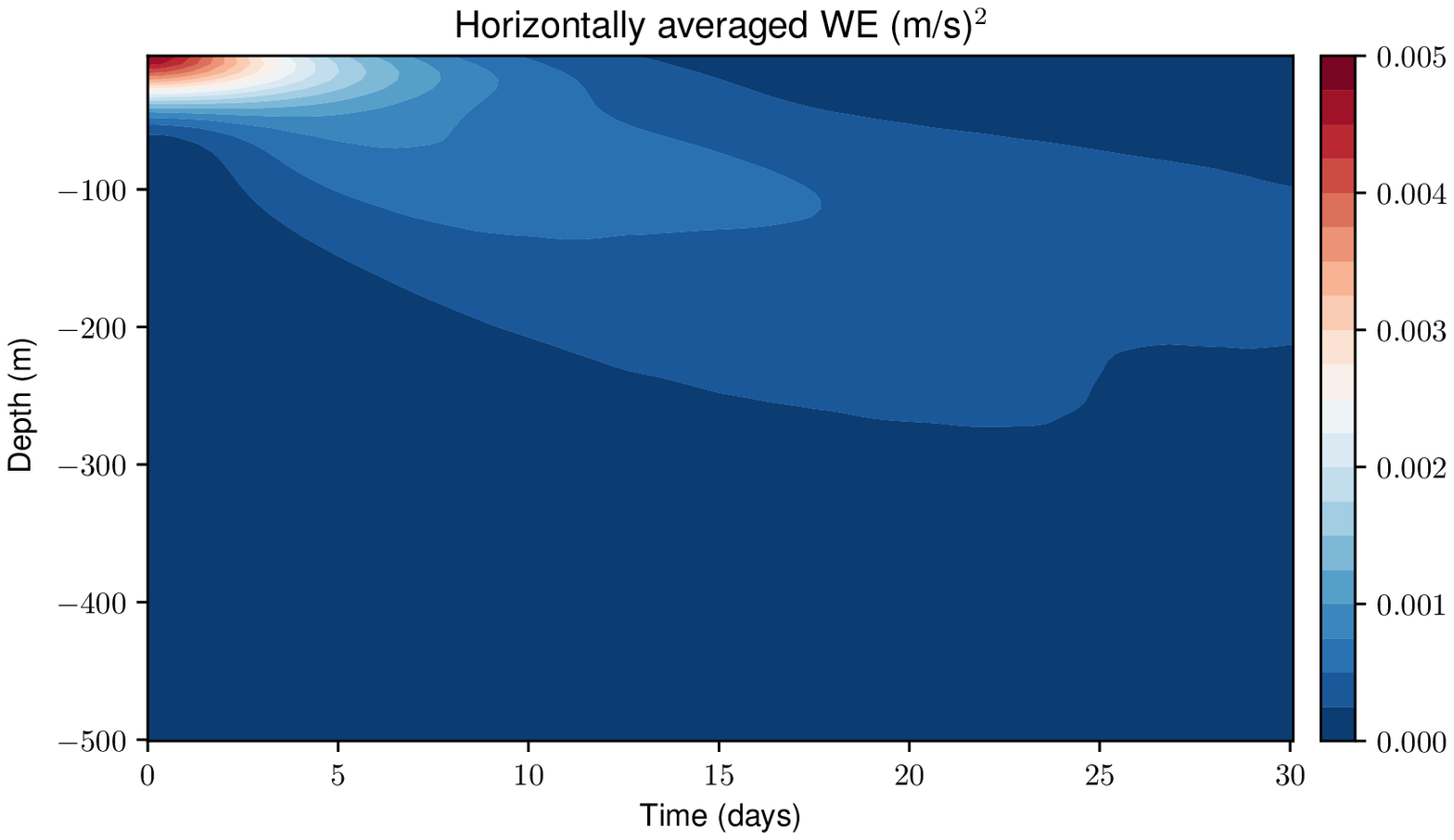}
\caption{Left: depth partition of total wave energy (WE) as a function of time. The white region is energy dissipated by horizontal hyperviscosity. Right: WE density distribution in the top 500 m.}
\label{wke_profile}
\end{figure*}

So far, we have seen that wave energy rapidly leaves the sea-surface (figure \ref{wke_zeta_long}), and localized vertical cuts revealed that at least some of this wave energy flows downwards through  anticyclonic drainpipes (figure \ref{wke_zeta_ver_shallow}). We conclude this section by quantifying the horizontally-averaged depth distribution of wave energy.

The left panel of figure \ref{wke_profile} shows the time evolution of the depth partition of wave energy. Following the wave initial condition, \eqref{waveIC}, over 90\% of WE is initially located in the upper 100 m. This energy is, however, rapidly drained to depth. After 20 days, the upper 100 m retains about only 20\% of the WE, as is typical in  observations \citep{dasaro1995,elipot}. Most of the energy is located in the top kilometer with about 5\% in each of the two bottom kilometers. 

Recall from \eqref{we_cons} that total wave energy is conserved in the absence of forcing and dissipative processes. In this initial value problem, loss of WE (top white region) may  thus only occur through diffusion operating at  small horizontal wave scales. After one month of evolution, about 20\% of wave energy is dissipated. This contrasts the minuscule $<1\%$ dissipative rates found in the wave escape scenario of \cite{cesar}, signaling that escape is less effective in a three-dimensional flow. This is not surprising: in the two-dimensional model of \cite{cesar} the vertical wavelength of waves is  fixed. This leads to an increase of group velocity as waves are deformed by the flow. Here the vertical wavelength is not fixed and wave capture \citep{bm2005} may occur. Over longer times scales, wave energy slowly radiates vertically. Equipartition between the top, middle and bottom kilometer bins necessitates about half a year of evolution  (not shown). By that time, about 70\% of WE is dissipated.


The right panel of figure \ref{wke_profile} provides a more detailed view of the WE vertical profile in the top 500 meters. Again, WE is rapidly drained out of the mixed layer, but vertical propagation stops at around 200 m, a depth comparable to that of the strongest vortices of the eddy field (see lower panels of figure \ref{IC}). This generalizes the intuition obtained from localized vertical cuts (figure \ref{wke_zeta_ver_shallow}): wave energy is drained down to the bottom of anticyclonic vortices. 


\section{Wave feedback} \label{sec:feed}

So far we have seen that eddies distort the primordial wind-generated inertial oscillation and guide wave energy into the ocean interior along anticyclonic drainpipes. In this simple scenario waves play a passive role: eddies are indifferent to the presence of waves. The QG-NIW model, however, includes a wave feedback term $q^w$ in \eqref{qeq}. Does this wave feedback mechanism play any role at all? 

\begin{figure*}[h!]
\centering
\includegraphics[trim=30 140 30 100, clip, width=.7\textwidth]{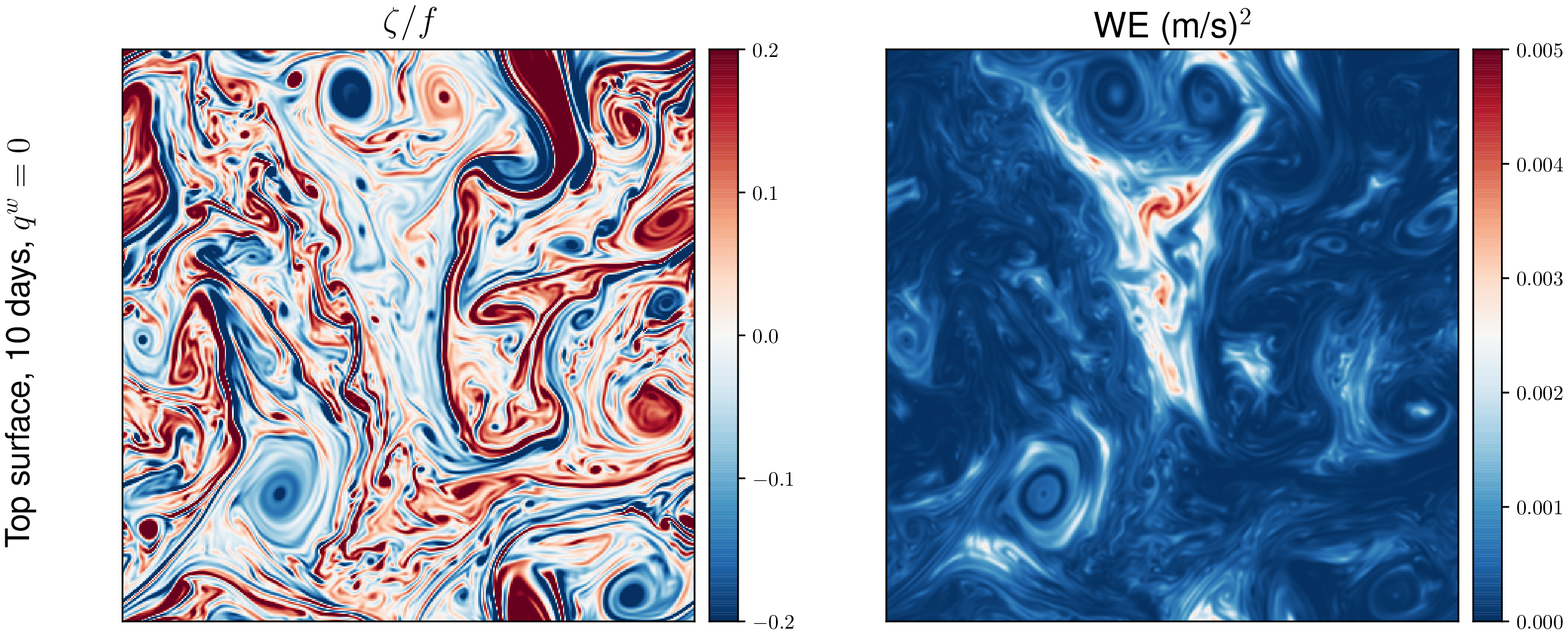}
\includegraphics[trim=30 140 30 140, clip, width=.7\textwidth]{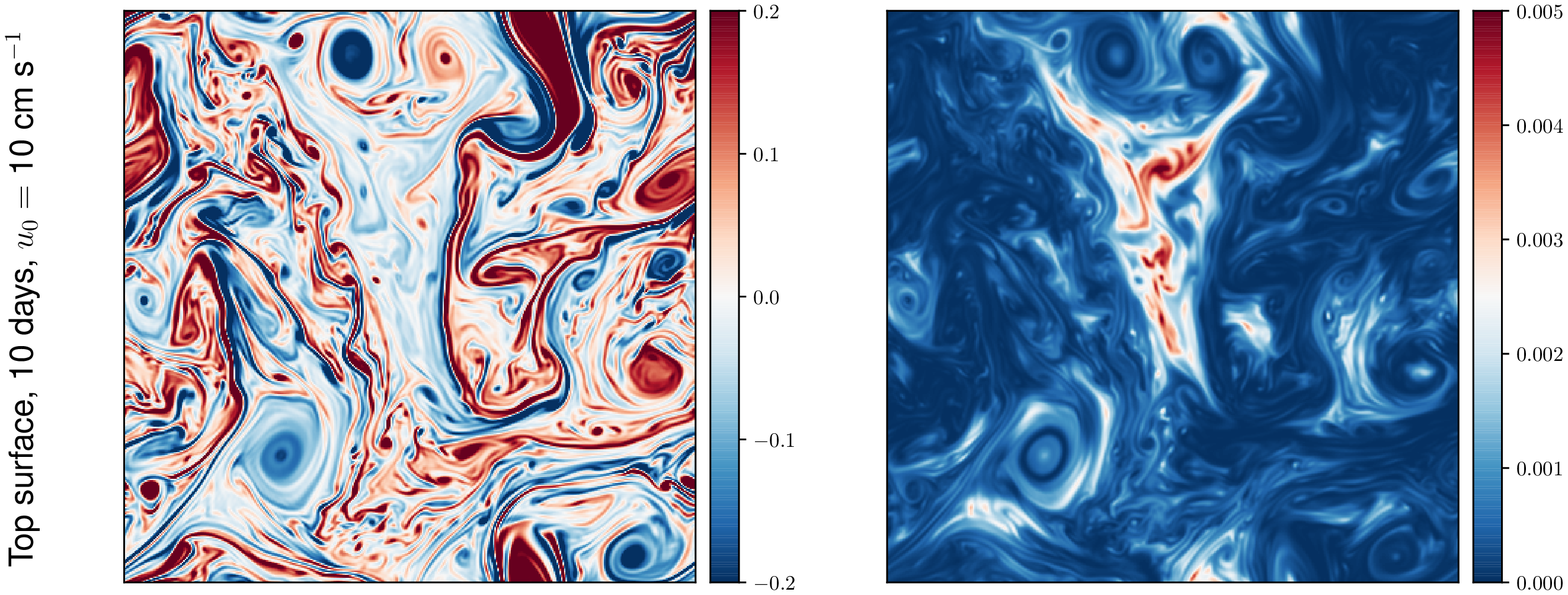}
\includegraphics[trim=30 140 30 140, clip, width=.7\textwidth]{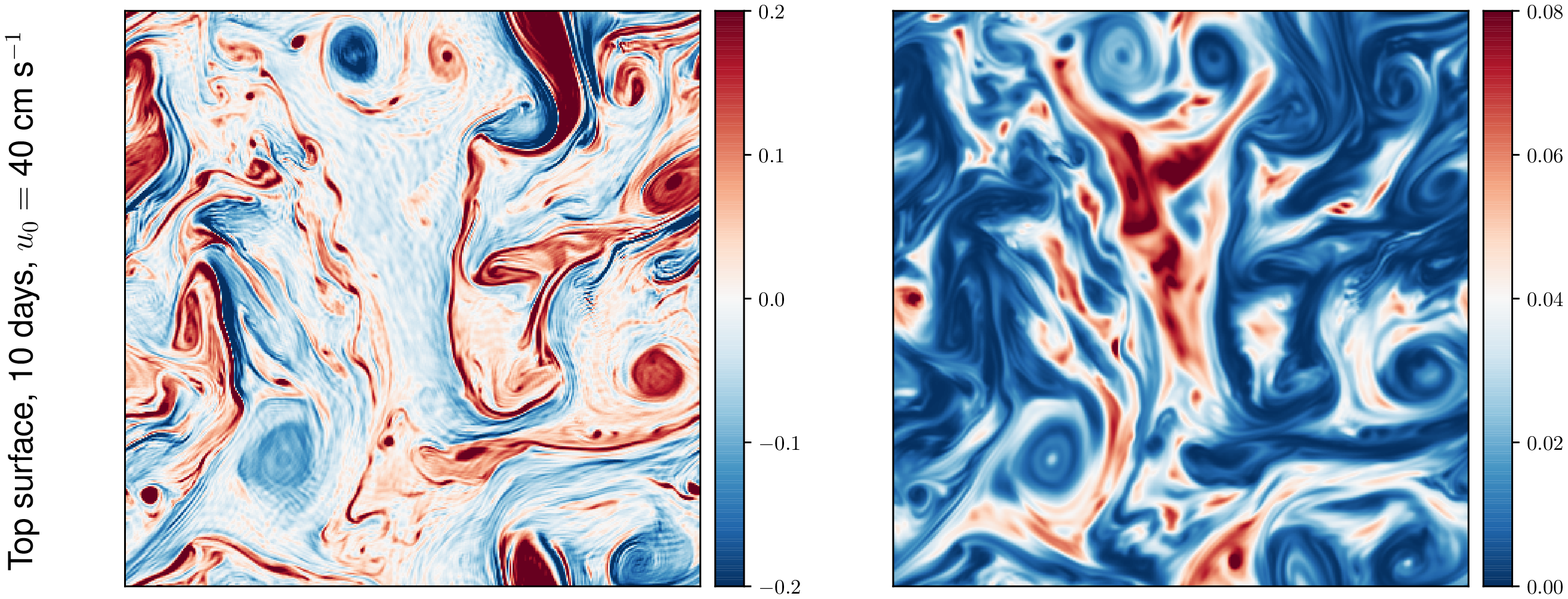}
\caption{Surface vorticity and WE after 10 days. Top: no feedback ($q^w = 0$, or equivalently, $u_0 = 0$ cm s$^{-1}$). Middle: standard storm with $u_0 = 10$ cm s$^{-1}$. Bottom: strong storm with $u_0 = 40$ cm s$^{-1}$. The wave energy colorbar is rescaled by a factor 16 for comparison. }
        \label{fb}
\end{figure*}

The answer, of course,  depends on the wind strength. In the preceding calculations we have taken the initial wave velocity $u_0 = 10$ cm s$^{-1}$. In this  case, although waves and eddies are equally strong, the wave feedback onto eddies is weak. The upper panels of figure \ref{fb} display surface $\zeta$ and WE in a non-interacting control run ($q^w=0$, or equivalently, $u_0=0$). Comparison with the middle panels, which shows the same fields but for the standard 10 cm s$^{-1}$ storm, indicates that wave feedback is negligible at the sea-surface.

Wave feedback is more apparent if the initial wave velocity  $u_0 = 10$ cm s$^{-1}$ is increased to $40$ cm s$^{-1}$ (bottom panels of figure \ref{fb}). Because $q^w$ is quadratic in the wave amplitude, this increases the wave feedback by a factor of 16. In this case wave feedback dampens vortices, especially smaller-scale cores and filaments. As a result, refractive imprinting is coarser. Wave scales are thus reduced more gently, which slows their penetration into the ocean interior. Thus  more wave energy remains at the surface after 10 days in the 40 cm s$^{-1}$ storm.

Wave feedback is also visible at the bottom of anticyclones, where wave energy collects and eddies weaken. Looking carefully at the vorticity cuts of figure \ref{wke_zeta_long10d}, which are from the standard 10 cm s$^{-1}$ run after 10 days, one may note the appearance of small-scale ``grooves'' with increasing depth. These grooves are collocated with accumulations of wave energy and are absent from simulations with $q^w=0$ (not shown).

\begin{figure*}
\centering
\includegraphics[trim=30 130 30 130, clip, width=.7\textwidth]{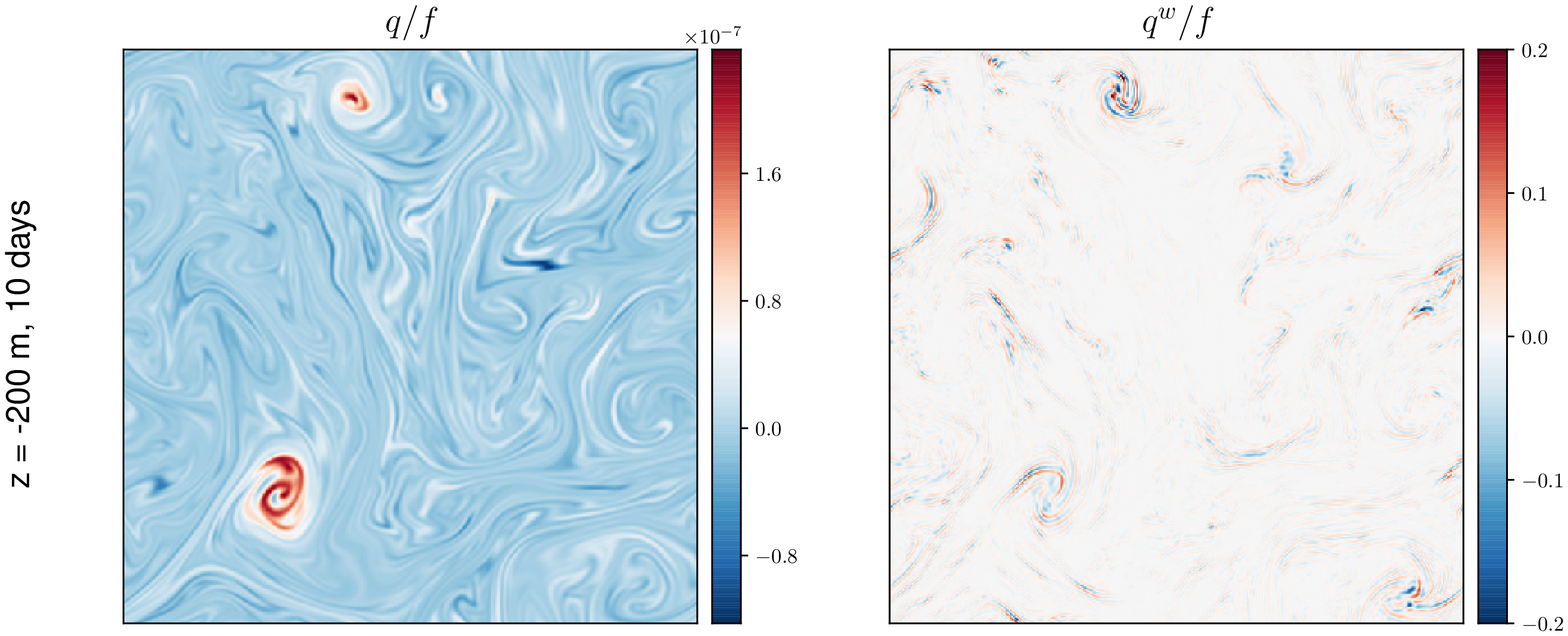}
\includegraphics[trim=30 130 30 130, clip, width=.7\textwidth]{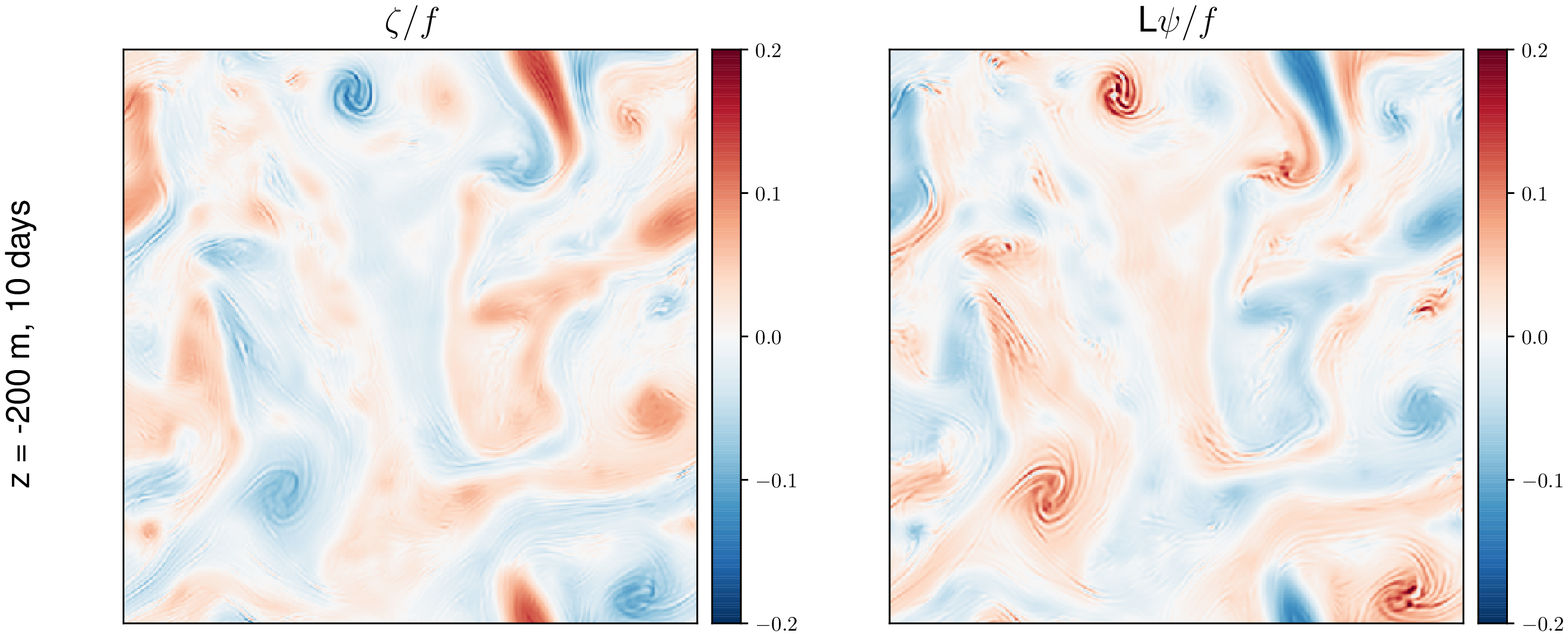}
\caption{Breakdown of potential vorticity, defined in \eqref{qeq}, for the standard run with $u_0 = 10$ cm s$^{-1}$, shown at $z=200$ m after 10 days. Note the colorbar for $q/f$ at left uses a different scale than the other three fields. }
\label{q_breakdown}
\end{figure*}

Figure \ref{q_breakdown} breaks down potential vorticity, $q$ defined in \eqref{qeq},  into  three constituents --- relative vorticity ($\zeta$), stretching ($\LL \psi$) and wave feedback ($q^w$) --- at the same time and location of the lowest panels of figure \ref{wke_zeta_long10d}. Note that $q$ is six orders of magnitude smaller than its three components: $q$ is the near-zero residual of the small perturbation used to seed the Eady spin-up (section \ref{sec:problem}). This signals there is near-perfect cancellation between the three components of $q$.

Also recall that the QG equation \eqref{qteq} expresses the conservation of potential vorticity, $q$. If $q$ is small and smooth initially --- as is the case here --- then it must remain so at all times. As the wave field develops horizontal gradients, $q^w$ becomes nonzero, and these changes must be compensated by relative vorticity and stretching to conserve the initial $q$. Because $q^w$ is comprised high-order derivatives of the wave field, it promotes the small-scale features of the wave field. It is the necessary cancellation of these small-scale $q^w$ features that produces grooves in the vorticity and stretching fields. 

These arguments are not limited to the Eady model, for which $q$ is specially zero. Even if $q$ is nonzero, QG dynamics dictates that $q$ is conserved such that the small-scale features in $q^w$ must be canceled by relative vorticity and stretching. A similar cancellation is seen in the turbulent simulations of \cite{cesar}, for which $q\neq 0$.

Despite these interesting features, wave feedback has, on the whole, a weak effect on wave propagation. When horizontal averages are performed, the depth partition of wave energy in figure \ref{wke_profile} is qualitatively unchanged, even for the vigorous 40 cm s$^{-1}$ storm (not shown). Sufficiently strong near-inertial waves do affect eddies through $q^w$, but not strongly enough to significantly alter the vertical propagation of the waves.

\section{Discussion} \label{sec:disc}

We have  explored the penetration of wind-generated near-inertial energy into a baroclinically turbulent ocean using a novel, two-way coupled three-dimensional QG-NIW model. Let's now weave our findings into a  narrative of the fate of near-inertial waves in the wake of a storm.

First, atmospheric storms with an appreciable inertial component (anticyclonic rotation near the local Coriolis frequency) sweep by the ocean, and the imparted momentum rapidly homogenizes in the mixed-layer, leading to the primordial slab-like inertial oscillation \citep{pollard_millard1970,dohandavis}. This initial wave has a horizontal scale comparable to that of the storm, and thus  its initial vertical propagation is glacially  slow \citep{gill1984}. During the first few days, however, $\zeta$-refraction leads to a collapse of the wave scales onto that of the vorticity field \citep{klein2004}. Wave energy then fluxes into anticyclonic regions \citep{DVB2015}. Anticyclones thereafter act as wave drains,  guiding wave energy and shear  downwards into the ocean interior \citep{leeniiler},  perhaps leading to mixing \citep{kunze1995}. These near-inertial drains terminate as the surface-intensified baroclinic vortices weaken at depth \citep{kunze1985}. Since the penetration depth of mesoscale structures is overall proportional to their horizontal scale \citep{lapeyresqg}, larger vortices penetrate more deeply  than smaller-scale filaments (figure \ref{IC}). Wave energy flows down these drain pipes and stalls where they terminate, so that horizontal cuts reveal wave energy distribution reminiscent of the terminating structures, with larger horizontal scales as depth is increased (figure \ref{wke_zeta_long10d}). 

Near-inertial waves play a mostly passive role in this narrative. Even for the most vigorous storm --- corresponding to surface wave energy an order of magnitude larger than that of the balanced flow  --- the effect on wave propagation is weak. Given the numerical burden imposed by the high-derivative $q^w$ term, the uncoupled QG-NIW model is a preferred option for efficiently examining wave propagation.

In the simulations described here, wind-generated near-inertial waves radiate out of the mixed-layer on the time scale of a week or two: see figure \ref{wke_profile}. After a month, the bulk ($\sim 60\%$) of wave energy is  between 100 and 1000m (corresponding to the termination depth of the  vortices) while 20\% has escaped to below 1km depth. We suspect, however,  that these simulations overestimate the amount of wave energy penetrating the ocean interior. In these computations  wave energy dissipates only  through horizontal diffusion. The QG-NIW model of section \ref{sec:model} does not represent shear instabilities unless a parametrization is supplied (not the case here). Shear instabilities would  provide a sink of wave energy via turbulent mixing \citep{lueck1986dissipation,kunze1995}. In fact, several investigators argue that a significant amount of near-inertial wave energy is lost to such turbulent mixing in the mixed layer \citep{zhai2009,jochum2013impact,soares2013}. 



\acknowledgments  This work was supported by the National Science Foundation Award  OCE-1657041 and by the Office of Naval Research award N00014-18-1-2803. We are grateful for computer resources provided by  the Extreme Science and Engineering Discovery Environment (XSEDE), which is supported by National Science Foundation grant number ACI-1548562. We thank Anna Savage for providing density profiles observed during the NISKINe pilot cruise. 



\bibliographystyle{ametsoc2014}
\bibliography{bibliography}



\end{document}